\documentclass[nopacs,nokeys]{revtex4}
\usepackage{amsfonts}
\usepackage{amsmath}
\usepackage{amssymb}
\usepackage{bbm}
\usepackage[utf8]{inputenc}
\usepackage{tensor}
\usepackage{slashed}
\usepackage[centertableaux ]{ytableau}
\usepackage{hyperref}
\usepackage{natbib}
\usepackage{enumerate}
\usepackage{braket,mleftright}
\usepackage{graphicx}
\usepackage[caption=false]{subfig}
\usepackage{subfig}
\usepackage{cleveref}

\begin{document}

\title{Bound states of the Yukawa potential from hidden supersymmetry}
 \author{ M. Napsuciale}
 \email{mauro@fisica.ugto.mx}
\address{Departamento de F\'{i}sica, Universidad de Guanajuato, Lomas del Campestre 103, Fraccionamiento
Lomas del Campestre, Le\'on, Guanajuato, M\'exico, 37150.}
\author{ S. Rodr\'{\i}guez } 
\email{simonrodriguez@uadec.edu.mx}
\address{Facultad de Ciencias F\'isico-Matem\'aticas,
  Universidad Aut\'onoma de Coahuila, Edificio A, Unidad
  Camporredondo, 25000, Saltillo, Coahuila, M\'exico.}

\begin{abstract}%
In this work, we present a phenomenological study of the complete analytical solution to the bound eigenstates and eigenvalues 
of the Yukawa potential obtained previously using the hidden supersymmetry of the system and a systematic expansion of the Yukawa 
potential in terms of $\delta=a_{0}/D$, where $a_{0}$ is the Bohr radius and $D$ is the screening length. 
The eigenvalues, $\epsilon_{nl}(\delta)$, are given in the form of Taylor series in $\delta$ which can be 
systematically calculated to the desired order $\delta^{k}$. 
Coulomb $l$-degeneracy is broken by the screening effects and, 
for a given $n$, $\epsilon_{nl}(\delta)$ is larger for higher values of $l$ which causes the crossing of levels for $n\ge4$. 
The convergence radius of the Taylor series can be enlarged up to the critical values using the Pad\'e approximants 
technique which allows us to calculate the eigenvalues with high precision in the whole rage of values of 
$\delta$ where bound states exist, and to reach a precise determination of the critical screening lengths, $\delta_{nl}$.
Eigenstates have a form similar to the solutions of the Coulomb potential, with the associated Laguerre polynomials replaced by new polynomials 
of order $\delta^{k}$ with $r$-dependent coefficients which, in turn, are polynomials in $r$. In general we find sizable deviations 
from the Coulomb radial probabilities only for screening lengths close to their critical values. We use these solutions to find the squared 
absolute value at the origin of the wave function for $l=0$, and their derivatives  for $l=1$, for the lowest states, as functions of $\delta$, 
which enter the phenomenology of dark matter bound states in dark gauge theories with a light dark mediator. 
\end{abstract}

\maketitle

\section{Introduction}
The Yukawa potential is the effective non-relativistic description of the interaction of two particles due to the exchange of a massive particle 
of mass $M$. It was proposed in Ref. \cite{Yukawa:1935xg} by H. Yukawa as a low energy description of the strong interactions between 
nucleons, due to the exchange of massive particles, now known as pions. The potential is given by
\begin{equation}
V(r)= - \alpha_{g} \frac{e^{-M r}}{r}, 
\label{YP}
\end{equation}
where $\alpha_{g}=g^{2}/4\pi$ denotes fine-structure constant of the interaction with coupling $g$. The range of the interaction is given 
by $D=1/M$, also named screening length. For $M=0$ we obtain the Coulomb potential and for large $M$ we have effective 
short-range interactions. The first estimate of the pion mass was done in \cite{Yukawa:1935xg} based on this potential and the first 
experimental results for the range of the nucleon-nucleon interactions.

The Yukawa potential appears in different branches of physics like plasma physics at low density and high temperatures 
\cite{Debye:1923sr}\cite{RevModPhys.31.569}\cite{PhysRev.125.1131}\cite{PhysRev.134.A1235}\cite{PhysRevA.27.418},
nuclear physics \cite{PhysRev.139.B1428}, astrophysics \cite{PhysRevA.9.52}
and solid state physics \cite{PhysRev.178.1337} \cite{PhysRevB.19.3167}\cite{FERRAZ1984627}. It is known as Debye-Huckel 
potential in plasma physics or Thomas-Fermi potential in solid state physics. In these applications, the interpretation of the 
effective parameters $\alpha_{g}$ and $M$ in Eq. (\ref{YP}) is different. In a cloud of charged ions and electrons at a 
temperature $T$, the Coulomb field produced by an ion is obtained from
the potential in Eq. (\ref{YP}) with $\alpha_{g}=Ze^{2}/4\pi$ and $D=1/M$ describes the Debye screening 
distance of the system given by $D=[k_{B}T/n_{e} e^{2}(1+Z^{2})]^{1/2}$, where $n_{e}$ is the electron density. For dopped 
semiconductors with injected carriers, $\alpha_{g}=e^{2}/4\pi \kappa$ where $\kappa$ stands for the dielectric
constant and $\lambda=1/M$ is the Thomas-Fermi screening length due to the injected carriers 
\cite{PhysRev.178.1337}  \cite{PhysRevB.19.3167}. 
  
The importance of the quantum Yukawa problem for many research areas of physics, motivated the search for solutions using diverse methods. 
It is well known that for a finite screening length there is a finite number of bound states \cite{PhysRev.134.A1235}\cite{PhysRev.139.B1428}
\cite{Bargmann961}. Approximate calculations for some of the energy levels exist, using variational methods \cite{PhysRev.125.1131}
 \cite{PhysRevA.9.52}\cite{PhysRev.178.1337}\cite{PhysRevA.8.1138}\cite{PhysRevA.48.220}\cite{PhysRevA.50.228}, Rayleigh-Schrodinger  
 perturbation theory \cite{PhysRev.134.A1235} \cite{PhysRevA.13.532}\cite{Edwards:2017ndv} \cite{PhysRevA.33.1433}
 \cite{PhysRevA.4.1875} \cite{Dutt_1985}; new perturbation schemes \cite{Eletsky:1981fm} \cite{Vainberg:1981}, and other methods
 \cite{PhysRevA.26.1759}\cite{PhysRevA.23.455}\cite{PhysRevLett.66.1310}\cite{PhysRevA.21.1100}\cite{Moreno:1983nc} 
 \cite{Gonul:2006}\cite{Patil_1984} \cite{PhysRevA.97.022503}. The ground state energy has been calculated to very high orders in the 
 expansion on the parameter $\delta =a_{0}/D=M/\alpha_{g}\mu$, where $a_{0}=1/\alpha_{g} \mu$ denotes the Bohr 
 radius and $\mu$ stands for the reduced mass of the bounded system. 

The problem of the bound states of the Yukawa potential has been also considered using numerical methods 
\cite{PhysRevA.1.1577}\cite{PhysRev.159.41}\cite{Diaz_1991}\cite{Li:2006chj}. Numerical results and available approximate analytical 
solutions for the lowest lying states shows that Coulomb $l$-degeneracy is broken, and for a given $n$, states with higher $l$ have a 
higher energy than lower $l$ states. Also, numerical solutions exhibit the phenomena of cross-over,  states 
of a given $n,l$ having a higher energy than states with $n+1,l^{\prime}$, which occurs close to the critical screening values (those for which 
a given state goes to the continuous) and are out of the reach of perturbative calculations. The value of the critical screening lengths are 
important for some applications, specially the ground state critical screening, and have been calculated using non-perturbative methods 
\cite{Patil_1984} and numerically solving the Yukawa potential for $n=0$ to $n=9$ \cite{PhysRevA.1.1577}. 

In the past few years, the Yukawa interaction has been also proposed as a solution to the core-cusp problem on the  
density profiles of dark matter in dwarf galaxies. The $N$-body simulations of collisionless cold dark matter predicts halo distributions 
singular at the center \cite{Navarro:2008kc} which is not observed in the data \cite{deBlok:2009sp}. Self-interacting dark matter has been 
proposed as a possibility to solve this problem \cite{Spergel:1999mh} but there is some tension with data 
\cite{Yoshida:2000bx}\cite{MiraldaEscude:2000qt}, which is alleviated if we consider that the self-interaction is induced by the exchange of 
a massive mediator which, for non-relativistic dark matter, yields a Yukawa potential \cite{Loeb:2010gj}\cite{Chan:2013yza}. 
The calculation of the dark matter profiles requires to solve the problem of classical scattering by the Yukawa potential in the strong 
coupling regime, a work done in \cite{Khrapak:2003kjw}. Yukawa interaction between two dark matter particles can also give rise to dark 
matter bound states (darkonium) which drives us to study of the Yukawa potential at the quantum level. 

We became interested in this problem in the search for gauge theory for spin-one dark matter (tensor dark matter in a spinor-like formalism)
\cite{Hernandez-Arellano:2018sen}\cite{Hernandez-Arellano:2019qgd}\cite{Napsuciale:2020kai}, whose simplest version is a 
hidden $U(1)_{DM}$ theory containing a new massive gauge boson $Z^{\prime}$. The formation of darkonium is an interesting possibility 
when the mediator can also couple to standard model particles. In the hidden scenario this can occur through the renormalizable kinetic 
mixing with the $U(1)_{Y}$ hypercharge \cite{Babu:1997st}\cite{Holdom:1985ag}\cite{Hewett:1988xc}\cite{Dienes:1996zr} 
\cite{Langacker:2008yv}.  On the experimental side, there are searches at the LHC for signatures of dark matter 
\cite{Aaboud:2017phn}\cite{Sirunyan:2017hci} looking for mono-jets events with a large missing energy which can be attributed to the 
production of a $Z^{\prime}$ which decay later into a pair of dark matter particle-antiparticle system. The possibility of the creation of 
darkonium has been studied recently and the corresponding phenomenology depends crucially on the bound state wave 
function or its derivatives at the origin 
\cite{MarchRussell:2008tu}\cite{An:2016gad}\cite{Cirelli:2016rnw}\cite{Petraki:2016cnz}\cite{Krovi:2018fdr} \cite{Harz:2017dlj}. The value 
of the squared wave function at the origin for the ground state, $|\psi (0)|^{2}$, has been estimated from variational methods 
\cite{PhysRevA.4.1875} and the obtained value used in phenomenological analysis \cite{Krovi:2018fdr}. 

In a previous letter, we proposed a procedure to solve this long standing problem \cite{Napsuciale:2020ehf}. Here, we give further 
details of the calculations and present a complete study of the phenomenology of the bound states of the Yukawa potential. The solution is 
based on the hidden supersymmetry of the Yukawa potential and on a perturbative expansion of the corresponding superpotentials in 
powers of $\delta $. We find that the quantum Yukawa problem is factorizable according to \cite{Infeld:1951mw}, up to order $\delta^{2}$. 
At the supersymmetry level this means that the supersymmetric partner of the Yukawa Hamiltonian $H_{l}$ belongs to the same family 
with different $l$ and we have "shape invariance" \cite{Gendenshtein:1984vs} of the effective potential $v_{l}(r)$. The complete spectrum 
can be obtained to this order following well known supersymmetric quantum mechanics techniques. Beyond ${\cal O}(\delta^{2})$ we 
loose shape invariance. However, hidden supersymmetry is still present and we can use it to completely solve the problem.
     
Our paper is organized as follows. In the next section we reduce the Yukawa potential to a unidimensional problem and review the basics 
of supersymmetric quantum mechanics in order to set our conventions. Section III is devoted to the solution of the Yukawa potential 
to leading order using supersymmetry and shape invariance and to set the main recurrence relations for the solution to higher orders. 
In section IV we use the same techniques to obtain the complete solution of the Yukawa potential up to ${\cal O} (\delta^{2})$. Section V is 
dedicated to obtain the complete analytical solution to  ${\cal O} (\delta^{3})$ constructing a family of supersymmetric 
Hamitonians. In section VI we generalize these results to arbitrary order in the expansion obtaining a systematic solution to the desired order.
Section VII is devoted to the analysis of the results for the main observables for the Yukawa potential. Our conclusions are given in section VIII 
and we close with an appendix with the explicit form of the polynomials entering the solution to order $\delta^{5}$ for $n=0,1,2,3,4$.

\section{Factorization and hidden supersymmetry of the Yukawa potential}

In the following we will work in conventional units recovering the $\hbar$ and $c$ factors. The radial Schrodinger equation for a particle of 
reduced mass $\mu$ in a central potential is
\begin{equation}
\left[-\frac{\hbar^{2}}{2\mu} \left(\frac{1}{r^{2}}\frac{d}{dr}(r^{2}\frac{d}{dr}) - \frac{l(l+1)}{r^{2}} \right)+V(r) \right]R(r)=E~R(r),
\end{equation}
where $l=0,1,2,..$ are the orbital angular momentum eigenvalues. Using the dimensionless variable $x=r/a$ where $a$ is a typical scale 
of the system it can be rewritten as
\begin{equation}
\left[-\frac{1}{x^{2}}\frac{d}{dx}(x^{2}\frac{d}{dx}) + \frac{l(l+1)}{x^{2}} +U(x) \right]R(ax)=\epsilon~R(ax),
\end{equation}
with the following shorthand notation
\begin{equation}
U(x)=\frac{2\mu a^{2}}{\hbar^{2}} V(ax), \qquad \epsilon=\frac{2\mu a^{2}}{\hbar^{2}} E.
\end{equation}
In terms of $ R(r)=u(x)/x$ this equation is reduced to the simple form
\begin{equation}
\left[-\frac{d^{2}}{dx^{2}} +v_{l}(x) \right]u_{l}(x)=\epsilon_{l}~u_{l}(x),
\label{eom}
\end{equation}
with the effective potential
\begin{equation}
v_{l}(x)\equiv \frac{l(l+1)}{x^{2}} +U(x).
\end{equation}
For the Yukawa potential in Eq.(\ref{YP}) we get
\begin{equation}
v_{l}(x)= \frac{l(l+1)}{x^{2}} -\frac{2}{x}e^{-\delta x},
\end{equation}
with the typical distance given by the Bohr radius of the system, $a=\frac{\hbar}{\mu c\alpha_{g}}\equiv a_{0}$ and we use the dimensionless ratio 
$\delta = a_{0}/D$, which in the following will be also named screening length. The energy levels can also be written in the simple form 
\begin{equation}
E_{l}=\frac{1}{2}\mu c^{2}\alpha^{2}_{g} \epsilon_{l}. 
\end{equation}

Now we factorize the Yukawa potential and construct the hidden supersymmetry of the system. First we write our problem as 
\begin{equation}
H_{l}u_{l}=\epsilon_{l}u_{l}
\end{equation}
where
\begin{equation}
H_{l}=-\frac{d^{2}}{dx^{2}} +v_{l}(x).
\end{equation}
This Hamiltonian can be factorized in terms of the following operators
\begin{equation}
a_{l}=-\frac{d}{dx}+W_{l}(x), \qquad a^{\dagger}_{l}=\frac{d}{dx}+W_{l}(x).
\end{equation}
A short calculation yields
\begin{align}
a_{l}a^{\dagger}_{l}&= -\frac{d^{2}}{dx^{2}} + W^{2}_{l}(x) - W^{\prime}_{l}(x),\\
a^{\dagger}_{l}a_{l}&= -\frac{d^{2}}{dx^{2}} + W^{2}_{l}(x) + W^{\prime}_{l}(x).
\end{align}
The factorization of $H_{l}$ in terms of these operators requires
\begin{equation}
W^{2}_{l}(x)- W^{\prime}_{l}(x)=v_{l}(x) - C(l,\delta),
\label{Riccati}
\end{equation}
in which case
\begin{equation}
H_{l} = a_{l} a^{\dagger}_{l} + C(l,\delta).
\end{equation}
The partner Hamiltonian is defined as
\begin{equation}
\tilde{H}_{l}=a^{\dagger}_{l} a_{l}+ C(l,\delta)=-\frac{d^{2}}{dx^{2}} +\tilde{v}_{l}(x),
\end{equation}
where
\begin{equation}
\tilde{v}_{l}(x)=  W^{2}_{l}(x)+ W^{\prime}_{l}(x) +C(l,\delta) = v_{l}(x) + 2 W^{\prime}_{l}(x).
\end{equation}

Now we construct the $2\times 2$ Hamiltonian 
\begin{equation}
H=\begin{pmatrix}a^{\dagger}_{l} a_{l}&0 \\ 0&a_{l} a^{\dagger}_{l} \end{pmatrix} 
=\left( -\frac{d^{2}}{dx^{2}} + W^{2}_{l}(x)\right) \mathbf{1}+ W^{\prime}_{l}(x) \sigma_{3},
\end{equation}
and charges
\begin{align}
Q_{1}&=\sigma_{1} p +\sigma_{2} W_{l}=\begin{pmatrix} 0& -i a^{\dagger}_{l} \\ i a_{l} & 0 \end{pmatrix},  \\
Q_{2}&= \sigma_{2} p -\sigma_{1} W_{l}= \begin{pmatrix} 0&  -a^{\dagger}_{l} \\ -a_{l} & 0 \end{pmatrix} ,
\end{align}
where $p=-i \frac{d}{dx}$. The charge operators satisfy
\begin{equation}
\{Q_{i},Q_{j}\} = 2 \delta_{ij} H, \qquad [Q_{i}, H]=0,
\end{equation}
which correspond to the $N=2$ supersymmetry (SUSY) algebra \cite{Witten:1981nf}\cite{COOPER1983262}.
The supersymmetric Hamiltonian is related to $H_{l}$ and $\tilde{H}_{l}$ as follows
\begin{equation}
H=\begin{pmatrix} \tilde{H}_{l} - C(l,\delta) & 0 \\ 0 & H_{l} -C(l,\delta) \end{pmatrix}
=\begin{pmatrix} -\frac{d^{2}}{dx^{2}} + U_{+}(x,l) & 0 \\0&  -\frac{d^{2}}{dx^{2}} + U_{-}(x,l)  \end{pmatrix} ,
\end{equation}
with the associated potentials
\begin{equation}
U_{\pm}(x,l)=W^{2}_{l}(x) \pm W^{\prime}_{l}(x).
\end{equation}
Notice that supersymmetry requires the super partner Hamiltonians to be intertwined by the $a_{l}$ and $a^{\dagger}_{l}$ 
operators, i.e.
\begin{equation}
a^{\dagger}_{l}H_{l}=\tilde{H}_{l}a^{\dagger}_{l}, \qquad a_{l}\tilde{H}_{l}=H_{l}a_{l}.
\end{equation} 
The group theoretical aspects of such relations have been studied in detail in \cite{Carinena:2001aj} and the realization
of the supersymmetric algebra in non-relativistic quantum mechanics with 
supercharges containing higher order derivatives, where similar intertwining relations and richer factorization 
schemes are obtained, have been developed in 
\cite{Andrianov:1993md} \cite{Andrianov:1994aj} \cite{FernandezC.:2005bs} \cite{Correa:2015wxa}. 

\section{Expansion and solution to leading order}
A complete analytical solution to Eq.(\ref{Riccati}) for the Yukawa potential
\begin{equation}
W^{2}_{l}(x,\delta)- W^{\prime}_{l}(x,\delta) + C(l,\delta)= \frac{l(l+1)}{x^{2}} -\frac{2}{x}e^{-\delta x} ,
\label{master}
\end{equation}
can be build, using an expansion of the superpotential in powers of $\delta$ \cite{Napsuciale:2020ehf}. 
First we decompose the superpotential into $\delta$-independent and $\delta$-dependent parts
\begin{equation}
W_{l}(x,\delta)=   w_{c}(x,l)+ w_{l}(x,\delta), \qquad C(l,\delta)=c(l)+ y(l,\delta).
\label{w}
\end{equation}
Next, we expand the effective potential in powers of $\delta$ 
\begin{equation}
v_{l}(x)=\frac{l(l+1)}{x^{2}} - \frac{2}{x} + 2 \delta  - \delta^{2} x + \frac{1}{3}\delta^{3} x^{2} + ... .
\label{veff}
\end{equation}
To leading order, Eq.(\ref{master}) involves only the $\delta$-independent part of the superpotential 
\begin{equation}
w^{2}_{c}(x,l)-w^{\prime}_{c}(x,l) + c(l)=\frac{l(l+1)}{x^{2}} - \frac{2}{x},
\label{wceq}
\end{equation} 
and the Yukawa problem reduces to the well known Coulomb problem. Writing the superpotential as 
\begin{equation}
w_{c}(x,l)=\frac{g(l)}{x} - \frac{1}{g(l)},
\label{Wdec}
\end{equation}
and inserting this expression in Eq. (\ref{wceq}) we find the conditions 
\begin{equation}
g(l)(g(l)+1)= l(l+1), \qquad  c(l)= -\frac{1}{g^{2}(l)}.
\end{equation}
There are two solutions for $g(l)$
\begin{equation}
g(l)=l ,\qquad  {\textrm{and}} \qquad g(l)=-(l+1).
\end{equation}
To ${\cal O}(\delta^{0})$ the partner Hamiltonian has the potential
\begin{equation}
\tilde{v}_{l}(x)= v_{l}(x) + 2 w^{\prime}_{c}(x,l)=\frac{l(l+1)-2g(l)}{x^{2}} - \frac{2}{x}.
\end{equation}
For $g(l)=l$ we get $\tilde{H}_{l}=H_{l-1}$ \cite{Infeld:1951mw}\cite{Fernandez:2000mv}. Similarly, for $g(l)=-(l+1)$ we obtain 
$\tilde{H}_{l}=H_{l+1}$.  We find it convenient to use $g(l)=-(l+1)$ and to work with the following super-potential for the Coulomb problem
\begin{equation}
 w_{c}(x,l)= \frac{1}{l+1}- \frac{l+1}{x},  \qquad  c(l) = -\frac{1}{(l+1)^{2}}.
 \label{wcoul}
\end{equation}
With this choice, the associated potentials satisfy 
\begin{align}
U_{+}(x,l)&\equiv w^{2}_{c}(x,l)+w^{\prime}_{c}(x,l)= v_{l+1}(x) - c(l), \\
U_{-}(x,l+1)&\equiv w^{2}_{c}(x,l+1)- w^{\prime}_{c}(x,l+1)= v_{l+1}(x) - c(l+1).
\end{align}
From these relations we can see that these potentials satisfy the "shape invariance" condition \cite{Gendenshtein:1984vs}
\begin{equation}
U_{+}(x,l)= U_{-}(x,f(l))+R(f(l)),
\end{equation}
where
\begin{equation}
f(l)=l+1, \qquad R(f(l))=c(f(l))-c(f(l-1)). 
\end{equation}
Following \cite{Gendenshtein:1984vs} we define $H^{0}(l)\equiv H_{l}$ and construct the family of Hamiltonians $\{ H^{r}(l) \}$,  defined by
\begin{equation}
H^{r}(l) =-\frac{d^{2}}{dx^{2}} + U_{-}(x,f^{(r)}(l)) + \sum^{r}_{i=1} R(f^{(i)}(l))
\label{Hr}
\end{equation}
with
\begin{equation}
f^{(i)}(l)=f(f^{(i-1)}(l))=f(f(f^{(i-2)}(l)))=...=l+i,
\end{equation}
where $f^{(i)}$ denotes the $i$-times composition of $f$. 
Using shape invariance we get 
\begin{align}
H^{r+1} (l)&=-\frac{d^{2}}{dx^{2}} + U_{-}(x,f(f^{(r)} (l)) +R(f(f^{(r)}(l))+ \sum^{r}_{i=1} R(f^{(i)}(l)) \nonumber \\
&=-\frac{d^{2}}{dx^{2}} + U_{+}(x,f^{(r)}(l)) + \sum^{r}_{i=1} R(f^{(i)}(l)),
\end{align}
thus, $H^{r}(l)-\sum^{r}_{i=1} R(f^{(i)}(l))$ and $H^{r+1}(l)-\sum^{r}_{i=1} R(f^{(i)}(l))$ are SUSY partners and have a common spectrum. 
In consequence, the whole family  $\{ H^{r}(l) \}$ has a common spectrum whose levels are given by
\begin{align}
\epsilon_{r,l}&=c(l) + \sum^{r}_{i=1} R(f^{(i)}(l))=-\frac{1}{(l+1)^{2}}+ \sum^{r}_{i=1} \left(\frac{1}{(l+i)^{2}}-\frac{1}{(l+i+1)^{2}}\right) 
\nonumber \\
&=-\frac{1}{(l+r+1)^{2}}.
\end{align}
In terms of the principal quantum number, $n=l+r+1$, we obtain 
\begin{equation}
\epsilon_{n,l}=-\frac{1}{n^{2}},
\end{equation}
where we changed the label $r\to n$. The angular momentum quantum number takes the values $l=n-1, n-2, ..., 1,0$ and all the 
corresponding states have the same energy.

The eigenstate $u_{n,l}(x)$ with the highest value of $l$ satisfies
\begin{equation}
H_{n-1}u^{c}_{n,n-1}(x)=\left( a_{n-1}a^{\dagger}_{n-1}+c(n-1)  \right) u^{c}_{n,n-1}(x) = \epsilon_{n,n-1}  u^{c}_{n,n-1}(x) .
\end{equation}
But from Eq. (\ref{wcoul}) we get
\begin{equation}
c(n-1)=-\frac{1}{n^{2}}=\epsilon_{n,n-1},
\end{equation}
thus, this state must satisfy $a^{\dagger}_{n-1}u^{c}_{n,n-1}=0$, a condition that can be used to obtain its explicit form. Indeed, 
  \begin{equation}
 \left[ \frac{d}{dx} +w_{c}(n-1,x) \right] u^{c}_{n,n-1}(x)=0,
  \end{equation}
  has the solution
  \begin{equation}
  u^{c}_{n,n-1}(x)=N_{n,n-1} e^{-\int w^{c}_{n-1}(x)dx}=N_{n,n-1} x^{n}e^{-\frac{x}{n}},
  \end{equation}
where $N_{n,n-1}$ is a normalization factor. For a given $n$, the eigenstates for lower values of $l$ can be obtained recursively 
from $u^{c}_{n,n-1}(x)$ using the lowering operator $a_{l}$
 \begin{align}
  u^{c}_{n,n-s}(x)&=N_{n,n-s} \left(-\frac{d}{dx}+ w_{c}(x,n-s)\right) u^{c}_{n,n-s+1}(x), 
  \end{align}
 where $s=2,3,...,n$. 
 
 This completes the solution of the Coulomb problem which is the leading term in the expansion of the Yukawa problem in powers of $\delta$.
 The explicit form of the wave functions is given in Section 6 where we address the general solution to the Yukawa potential and with a 
 convenient choice in the phases of the normalization factors we recover the conventional solutions to the Coulomb problem in terms of 
 the Laguerre associated polynomials.
 
 Once solved the leading order, we use the decomposition (\ref{w}) in Eq.(\ref{master}), to get the following equation for the 
 complementary function $w_{l}(x,\delta)$
\begin{align}
w^{2}_{l}(x,\delta)- w^{\prime}_{l}(x,\delta) + 2 w_{c}(x,l) w_{l}(x,\delta) &=  2 \delta  - \delta^{2} x + \frac{1}{3}\delta^{3} x^{2} +... - y(l,\delta).
\label{weq}
\end{align}
Notice that the expansion in the right hand side (r.h.s.) of this equation starts at order $\delta$ and it is also an expansion in powers of $x$. 
The ${\cal O}(\delta^{k})$ term on the r.h.s. is ${\cal O}(x^{k-1})$. There is always a polynomial solution in $x$ with 
$\delta$-dependent coefficients for $w_{l}(x,\delta)$ when we work to ${\cal O}(\delta^{k})$, with the advantage that powers of $\delta$ and 
$x$ are correlated. Indeed, the general solution can be written as
\begin{align}
w_{l}(x,\delta)&=a_{1}\delta+ (a_{2}\delta^{2}+a_{3}\delta^{3}+a_{4}\delta^{4}...) x + (b_{3}\delta^{3} + b_{4}\delta^{4}+b_{5}\delta^{5}+...) x^{2} 
\nonumber \\
&+ ( c_{4}\delta^{4} + c_{5}\delta^{5}+c_{6}\delta^{6} +...) x^{3} +..., \\
y(l,\delta)&=y_{1}(l) \delta + y_{2}(l) \delta^{2} +  y_{3}(l) \delta^{3} + ... \label{Lexp}
\end{align}
The coefficients required in the calculation to a given order in $\delta$ can be fixed matching powers of $x$ on both sides of Eq. (\ref{weq}).

\section{Solution of the Yukawa problem to order $\delta^{2}$ }

The solution to order ${\cal O}(\delta)$ can be written as
\begin{equation}
 w_{l}(x,\delta)=a_{1}\delta , \qquad y(l,\delta)=y_{1}(l) \delta. 
\end{equation}
Inserting these expressions in Eq. (\ref{weq}) and comparing powers of $x$ we find the solution $a_{1}=0$, $y_{1}(l)=2$, thus the solution to this
order is given by
\begin{equation}
{\cal O}(\delta):\qquad w_{l}(x,\delta)=0, \qquad y(l,\delta)=2\delta.
\end{equation}
To  ${\cal O}(\delta^{2})$ the solution must be of the form
\begin{equation}
 w_{l}(x,\delta)=a_{2} \delta^{2} x, \qquad y(l,\delta)= 2\delta + y_{2}(l)\delta^{2}.
\end{equation}
Inserting these relations in Eq.(\ref{weq}), matching powers of $x$ and keeping only up to ${\cal O}(\delta^{2})$ terms, we obtain the 
unique solution
$a_{2}=-(l+1)/2$ and $y_{2}(l)=- (l+1)(l+3/2)$, thus the solution to this order is
\begin{equation}
{\cal O}(\delta^{2}):\qquad w_{l}(x,\delta)=-\frac{1}{2}(l+1) \delta^{2} x, \qquad y(l,\delta)=  2\delta - (l+1)(l+\frac{3}{2})\delta^{2}.
\end{equation}

 \subsection{Energy levels at ${\cal O}(\delta^{2})$}
The solution to ${\cal O}(\delta)$ is straightforward since in this case we just add a constant  $2\delta$ to the Coulomb potential, thus 
we get the Coulomb eigenstates with the corresponding energy levels shifted by $2\delta$. 

The first non-trivial case appears at ${\cal O}(\delta^{2})$. In this case the solution is
 \begin{equation}
W_{l}(x,\delta)= - \frac{1}{l+1} - \frac{l+1}{x}  - \frac{1}{2}(l+1)\delta^{2}x, \qquad 
C(l,\delta)=-\frac{1}{(l+1)^{2}}+2\delta -(l+1)(l+\frac{3}{2})\delta^{2}.
\label{Wl2Cl2}
\end{equation}
The Hamiltonian is factorized as
\begin{equation}
H_{l} =-\frac{d^{2}}{dx^{2}}+ \frac{l(l+1)}{x^{2}}-\frac{2}{x}+2\delta-\delta^{2} x = a_{l}a^{\dagger}_{l}+ C(l,\delta).
\end{equation}
The partner Hamiltonian reads
\begin{equation}
\tilde{H}_{l}=a^{\dagger}_{l}a_{l}+C(l,\delta)=-\frac{d^{2}}{dx^{2}}+\frac{(l+1)(l+2)}{x^{2}}-\frac{2}{x}+2\delta -\delta^{2} x - (l+1)\delta^{2},
\end{equation}
such that
\begin{equation}
\tilde{H}_{l-1}=a^{\dagger}_{l-1}a_{l-1}+C(l-1,\delta)=-\frac{d^{2}}{dx^{2}}+\frac{l(l+1)}{x^{2}}-\frac{2}{x} + 2\delta-\delta^{2} x - l\delta^{2}
=H_{l} - l\delta^{2}.
\end{equation} 
 The partner Hamiltonian is related to the original one although a new ($\delta$-dependent) constant term is generated
 \begin{align}
a_{l} a^{\dagger}_{l} &=H_{l}-C(l,\delta), \label{Hada}\\
a^{\dagger}_{l-1}a_{l-1}&=H_{l}-C(l-1,\delta) - l\delta^{2}= H_{l}-\tilde{C}(l-1,\delta) ,
\label{Haad}
\end{align} 
 where
 \begin{equation}
\tilde{C}(l,\delta)=C(l,\delta)+(l+1)\delta^{2}.
 \end{equation}
 In this case, the associated potentials satisfy 
\begin{align}
U_{+}(x,l)&\equiv W^{2}_{l}(x,\delta)+W^{\prime}_{l}(x,\delta)= v_{l+1}(x,\delta) - \tilde{C}(l,\delta), \\
U_{-}(x,l+1)&\equiv W^{2}_{l+1}(x,\delta)-W^{\prime}_{l+1}(x,\delta)= v_{l+1}(x,\delta) - C(l+1,\delta),
\end{align}
and still satisfy the "shape invariance" condition
\begin{equation}
U_{+}(x,l)= U_{-}(x,f(l))+R(f(l))
\end{equation}
with
\begin{equation}
f(l)=l+1, \qquad R(f(l))=C(f(l),\delta )-\tilde{C}(f(l-1),\delta ). 
\end{equation}
Now we follow a similar procedure to the Coulomb case defining the analogous family $\{ H^{r}\}$ of SUSY partners in Eq.(\ref{Hr}) to obtain 
the spectrum of $H^{0}=H_{l}$ as 
\begin{align}
\epsilon_{r,l}&=C(l,\delta) + \sum^{r}_{i=1} R(f^{(i)}(l)) \nonumber \\
&=-\frac{1}{(l+r+1)^{2}} +2\delta - \left[(l+1)(l+\frac{3}{2}) +3r(r+2(l+1))\right]\delta^{2},
\end{align}
which when written in terms of the principal quantum number, $n=l+r+1$, reads
\begin{equation}
\epsilon_{n,l}=-\frac{1}{n^{2}}+2\delta - \frac{1}{2}[3n^{2}-l(l+1)]\delta^{2}.
\label{E2nl}
\end{equation}

 \subsection{Eigenstates to ${\cal O}(\delta^{2})$}

The eigenstate $u_{n,l}(x)$ for $l=n-1$ satisfies
\begin{equation}
H_{n-1}u_{n,n-1}(x)=\left( a_{n-1}a^{\dagger}_{n-1}+C(n-1,\delta)  \right) u_{n,n-1}(x) = \epsilon_{n,n-1}  u_{n,n-1}(x) .
\end{equation}
Comparing Eqs.(\ref{Wl2Cl2},\ref{E2nl}) we obtain
\begin{equation}
C(n-1,\delta)=-\frac{1}{n^{2}}+2\delta-n(n+\frac{1}{2})=\epsilon_{n,n-1}.
\end{equation}
This state must satisfy $a^{\dagger}_{n-1}u_{n,n-1}=0$, i.e.  
\begin{equation}
 \left[ \frac{d}{dx} +W_{n-1}(x,\delta) \right] u_{n,n-1}(x)=0,
  \end{equation}
  which can be solved to obtain 
  \begin{equation}
  u_{n,n-1}(x,\delta)=N_{n,n-1}(\delta ) e^{-\int W_{n-1}(x,\delta)dx}=N_{n,n-1}(\delta ) x^{n}e^{-\frac{x}{n} +\frac{1}{4}n\delta^{2}x^{2}}.
  \label{sol2nm1}
  \end{equation}

Using Eqs.(\ref{Hada},\ref{Haad}) we obtain recursively states with lower $l$ acting with the operator $a_{l}$
 \begin{align}
  u_{n,n-s}(x)&=N_{n,n-s} (\delta) ~ a_{n-s}u_{n,n-s+1}.
  \label{sol2nmk}
  \end{align}
 
 This yields the complete solution of the quantum Yukawa problem to order $\delta^{2}$. We remark that actually the systematic calculation 
 of the eigenstates to order $\delta^{2}$ requires the expansion of both the exponential and the normalization factors in 
 Eqs. (\ref{sol2nm1},\ref{sol2nmk}). We will address the details of this expansion in the general case considered in Section 6, 
 but it is clear that this procedure will replace these factors by polynomials of order $\delta^{2}$ with $x$-dependent coefficients.
 
\section{Solution to the Yukawa potential to ${\cal O}(\delta^{3})$}
 To  ${\cal O}(\delta^{3})$ the superpotential is given by 
 \begin{equation}
W_{l}(x)=    \frac{1}{l+1} - \frac{l+1}{x} -\left[\frac{1}{2}(l+1) \delta^{2}+\frac{1}{6}(l+1)^{2}(l+2)\delta^{3}\right] x + \frac{l+1}{6}\delta^{3} x^{2}.
\end{equation}
The Hamiltonian is factorized as
\begin{equation}
H_{l} =-\frac{d^{2}}{dx^{2}}+ \frac{l(l+1)}{x^{2}}-\frac{2}{x}+2\delta-\delta^{2} x +\frac{1}{3}\delta^{3}x^{2}= a_{l}a^{\dagger}_{l}+ C(l,\delta).
\end{equation}
where
\begin{equation}
C(l,\delta)=-\frac{1}{(l+1)^{2}}+ 2\delta - (l+1)(l+\frac{3}{2})\delta^{2} +\frac{1}{3} (l+1)^{2}(l+2)(l+\frac{3}{2})\delta^{3}.
\end{equation}
The partner Hamiltonian now is given by
\begin{equation}
\tilde{H}_{l-1}=a^{\dagger}_{l-1}a_{l-1}+C(l-1,\delta) =H_{l} - l\delta^{2} +\frac{1}{3}l[l(l+1) + 2x]\delta^{3}.
\end{equation}
 At ${\cal O}(\delta^{3})$, the Hamiltonians $H_{l}$ and $H_{l-1}$ are not longer connected by supersymmetry and we loose the shape 
invariance of the potentials of the family $\{ H_{l} \}$ which allowed us to solve easily the problem at lower orders. Notice however that 
shape invariance is not a necessary condition for supersymmetry. Indeed, shape invariance holds when we have supersymmetry {\it and } 
the Hamiltonian family is {\it factorizable} as defined in \cite{Infeld:1951mw}. This is not the case for the Yukawa potential beyond 
${\cal O}(\delta^{2})$ but the hidden supersymmetry of the system still holds and we will use it to find a complete analytical solution 
to the problem. First, following the pattern obtained at lower orders, we expect that 
the state with the highest value of $l$ still satisfy the condition $a^{\dagger}_{n-1} u_{n,n-1}=0$ which yields
  \begin{equation}
  u_{n,n-1}(x,\delta)=N_{n,n-1}(\delta ) e^{-\int W_{n-1}(x,\delta)dx}=N_{n,n-1}(\delta ) x^{n} e^{-\frac{x}{n}}
  e^{[\frac{n}{2}\delta^{2}-\frac{n}{6}(n+1)\delta^{3}]\frac{x^{2}}{2} - \frac{n}{6} \delta^{3}  \frac{x^{3}}{3}  }. 
  \end{equation}
A direct calculation shows that this function indeed satisfies Eq. (\ref{eom}) with eigenvalue
\begin{equation}
\epsilon_{n,n-1}=-\frac{1}{n^{2}}+2\delta-n(n+\frac{1}{2})\delta^{2} + \frac{1}{3}n^{2}(n+1)(n+\frac{1}{2})\delta^{3}=C(n-1,\delta).
\end{equation}
The state $u_{n,n-2}(x,\delta)$ is no longer obtained as $a_{n-2}u_{n,n-1}(x,\delta)$ as a direct calculation shows. This is due to 
the fact that $H_{l}$ and $H_{l-1}$ are not longer connected by supersymmetry. However, $H_{l}$ and $\tilde{H}_{l}$ are SUSY partners,
thus they are isospectral, and we can try to solve $\tilde{H}_{l}$ in whose case we would be finding also the eigenvalues of $H_{l}$. 
With this aim, let us take a closer look to the SUSY partner of $H_{l}$. This will be the first step towards the complete solution 
to this order, thus we denote the partner as $\tilde{H}^{(1)}_{l}$ in the following. It is given by
\begin{equation}
\tilde{H}^{(1)}_{l}\equiv a^{\dagger}_{l}a_{l}+C(l,\delta)=-\frac{d^{2}}{dx^{2}}+U_{+}(x,l)+C(l,\delta)=\frac{d^{2}}{dx^{2}}+\tilde{v}^{(1)}_{l}(x),
\end{equation}
where to  ${\cal O}(\delta^{3})$ 
\begin{equation}
\tilde{v}^{(1)}_{l}(x)=v_{l+1}(x)-(l+1)\delta^{2}+\frac{1}{3}(l+1)[(l+1)(l+2)+2x]\delta^{3}.
\end{equation}
Since $H_{l}$ and $\tilde{H}^{(1)}_{l}$ are SUSY partners they must have a common spectrum. We can solve 
$\tilde{H}^{(1)}_{l}$ at least for the highest allowed value of $l$ following the same procedure used to solve $H_{l}$ for $l=n-1$. 
First we re-factorize $\tilde{H}^{(1)}_{l}$ as follows
\begin{equation}
\tilde{H}^{(1)}_{l}=\tilde{a}_{l}^{(1)} (\tilde{a}_{l}^{(1)})^{\dagger}+\tilde{C}^{(1)}(l,\delta)
\end{equation}
where
\begin{align}
\tilde{a}_{l}^{(1)}=-\frac{d}{dx}+\tilde{W}_{l}^{(1)}(x), \qquad (\tilde{a}_{l}^{(1)})^{\dagger}=\frac{d}{dx}+\tilde{W}^{(1)}_{l}(x).
\end{align}
The new superpotential $\tilde{W}^{(1)}_{l}(x)$ is decomposed as
\begin{equation}
\tilde{W}^{(1)}_{l}(x,\delta)= w_{c}(x,l+1) + \tilde{w}^{(1)}_{l}(x,\delta),
\end{equation}
and $ \tilde{w}^{(1)}_{l}(x,\delta)$ must satisfy an equation similar to Eq.(\ref{weq}), but with the corresponding expansion of $\tilde{v}^{(1)}_{l}$ 
on the right hand side, namely 
 \begin{align}
(\tilde{w}^{(1)}_{l}(x,\delta))^{2}&- (\tilde{w}^{(1)}_{l}(x,\delta))^{\prime} + 2 w_{c}(x,l+1) \tilde{w}^{(1)}_{l}(x,\delta) =  \nonumber \\
 & 2 \delta  -(l+1) \delta^{2}  + \frac{1}{3}(l+1)^{2}(l+2)\delta^{3}
  - (\delta^{2}-\frac{2}{3}(l+1)\delta^{3})x +\frac{1}{3}\delta^{3} x^{2}.
\label{tweq}
\end{align}
This equation can be used to obtain $\tilde{w}^{(1)}_{l}(x,\delta)$ and $\tilde{C}^{(1)}(l,\delta)$ to ${\cal O}(\delta^{3})$ as we did 
for $w_{l}(x,\delta)$. We get
\begin{align}
\tilde{w}^{(1)}_{l}(x,\delta)&=\left[ -\frac{1}{2}(l+2)\delta^{2}+\frac{1}{6}(l+2)(l^{2}+7 l + 8)\delta^{3}\right] x +\frac{1}{6}(l+2)\delta^{3} x^{2}, \\
\tilde{C}^{(1)}(l,\delta)&= -\frac{1}{(l+2)^{2} }+ 2\delta -(l+\frac{3}{2})(l+4)\delta^{2} +\frac{1}{3}(l+\frac{3}{2})(l+2)^{2}(l+7)\delta^{3}.
\end{align}
The solution of this potential for $l=n-2$ is
\begin{equation}
\tilde{u}^{(1)}_{n,n-2}(x)= e^{\int \tilde{W}^{(1)}_{n-2}(x) dx}=
x^{n}e^{-\frac{x}{n}} e^{\frac{1}{2}n\delta^{2}-\frac{1}{12}n (n^{2}+3 n - 2)\delta^{3} x^{2}- \frac{1}{18}n\delta^{3} x^{3}},
\end{equation} 
and the corresponding energy is
\begin{equation}
\tilde{\epsilon}^{(1)}_{n,n-2}=\tilde{C}^{(1)}(n-2,\delta)=-\frac{1}{n^{2} }+ 2\delta -(n-\frac{1}{2})(n+2)\delta^{2} +\frac{1}{3}(n-\frac{1}{2})n^{2}(n+5)\delta^{3}.
\end{equation}
 Now we can find the eigenstate of $H_{l}$ for $l=n-2$ using the double factorization 
  \begin{align}
 \tilde{H}^{(1)}_{n-2}=\tilde{a}^{(1)}_{n-2}(\tilde{a}^{(1)}_{n-2})^{\dagger}+\tilde{C}^{(1)}(n-2,\delta) =a^{\dagger}_{n-2} a_{n-2}+C(n-2,\delta).
 \end{align}
 The state $\tilde{u}^{(1)}_{n,n-2}(x)$ satisfies
 \begin{align}
 [a^{\dagger}_{n-2} a_{n-2}+C(n-2,\delta)]\tilde{u}^{(1)}_{n,n-2}=\tilde{\epsilon}^{(1)}_{n,n-2}\tilde{u}^{(1)}_{n,n-2}.
 \end{align}
 Acting on the last equation with $a_{n-2}$ we get
  \begin{align}
H_{n-2}(a_{n-2}\tilde{u}_{n,n-2})=[a_{n-2}a^{\dagger}_{n-2} +C(n-2,\delta)](a_{n-2}\tilde{u}_{n,n-2})
=\tilde{\epsilon}^{(1)}_{n,n-2}(a_{n-2}\tilde{u}^{(1)}_{n,n-2}),  \end{align}
 and we obtain the same energy level for $H_{n-2}$ and $\tilde{H}^{(1)}_{n-2}$ as expected
 \begin{equation}
\epsilon_{n,n-2}=\tilde{\epsilon}^{(1)}_{n,n-2}=-\frac{1}{n^{2} }+ 2\delta -(n-\frac{1}{2})(n+2)\delta^{2} +\frac{1}{3}(n-\frac{1}{2})n^{2}(n+5)\delta^{3},
 \end{equation}
while the eigenstate is given by
\begin{equation}
u_{n,n-2}=N_{n,n-2} (\delta ) \, a_{n-2}\tilde{u}^{(1)}_{n,n-2}.
\end{equation}
 
 Eigenstates and eigenvalues for $l=n-3$ can be calculated applying now this procedure to $\tilde{H}^{(1)}_{l}$. The 
 SUSY partner, denoted $\tilde{H}^{(2)}_{l}$ is
 \begin{equation}
 \tilde{H}^{(2)}_{l}\equiv (\tilde{a}^{(1)}_{l})^{\dagger}\tilde{a}^{(1)}_{l} + \tilde{C}^{(1)}(l,\delta)\equiv \frac{d^{2}}{dx^{2}}+\tilde{v}^{(2)}_{l}(x).
 \label{H2l}
 \end{equation}
 We re-factorize this Hamiltonian as
 \begin{equation}
 \tilde{H}^{(2)}_{l}\equiv \tilde{a}^{(2)}_{l} (\tilde{a}^{(2)}_{l})^{\dagger}+ \tilde{C}^{(2)}(l,\delta),
 \label{H2lbis}
 \end{equation}
 with
 \begin{equation}
 \tilde{a}^{(2)}_{l} =-\frac{d}{dx}+\tilde{W}^{(2)}_{l}(x), \qquad   (\tilde{a}^{(2)}_{l})^{\dagger}= \frac{d}{dx}+\tilde{W}^{(2)}_{l}(x).
 \end{equation}
Following the same procedure we obtain
\begin{align}
\tilde{v}^{(2)}_{l}(x)&=v_{c}(l+2)+2\delta -(2l+3)\delta^{2}+\frac{2}{3}(l+\frac{3}{2}) (l+2)(l+3) \nonumber \\
&- (\delta^{2}-\frac{4}{3}(l+\frac{3}{2})\delta^{3})x 
+ \frac{1}{3}\delta^{3}x^{2}, \\
\tilde{W}^{(2)}_{l}(x)&=w_{c}(l+2)+ \left[ -\frac{1}{2}(l+3)\delta^{2} + \frac{1}{6}(l+2)(l+3)(l+9)\delta^{3}\right] x +\frac{1}{6}(l+3)\delta^{3}x^{2}, \\
 \tilde{C}^{(2)}(l,\delta)&= -\frac{1}{(l+3)^{2}} +2\delta - \frac{1}{2} (2l^{2}+17l+27)\delta^{2} +\frac{1}{3} (l+3)^{2}(l+2)(l+\frac{23}{2}) \delta^{3}.
\end{align}
The common eigenvalue of the three Hamiltonians $\{H_{n-3},\tilde{H}^{(1)}_{n-3},\tilde{H}^{(2)}_{n-3} \}$ is
\begin{equation}
\epsilon_{n,n-3}=-\frac{1}{n^{2}} + 2\delta - (n^{2}+\frac{5}{2}n-3)\delta^{2} +\frac{1}{3} n^{2}(n-1)(n+\frac{17}{2}) \delta^{3}.
\end{equation}
The corresponding eigenstates are given by
\begin{align}
\tilde{u}^{(2)}_{n,n-3}(x)&= N^{(2)}_{n,n-3} (\delta ) e^{\int \tilde{W}^{(3)}_{n-3}(x) dx}, \label{sol3nm3}\\
\tilde{u}^{(1)}_{n,n-3}(x)&=N^{(1)}_{n,n-3}(\delta ) \tilde{a}^{(1)}_{n-3}\tilde{u}^{(2)}_{n,n-3} (x),\\
u_{n,n-3}(x)&=N^{(0)}_{n,n-3} (\delta ) a_{n-3}\tilde{u}^{(1)}_{n,n-3}(x)=N_{n,n-3}(\delta ) a_{n-3}\tilde{a}^{(1)}_{n-3}\tilde{u}^{(2)}_{n,n-3}(x).
\end{align}
 Continuing this process we will eventually reach the lowest $l=0$ level, completely solving the Yukawa problem to order ${\cal O}(\delta^{3}) $. 
 The complete set of eigenvalues to ${\cal O}(\delta^{3}) $ is given by 
 \begin{equation}
  \epsilon_{n,l}(\delta) =  - \frac{1}{n^{2}} + 2\delta - \frac{1}{2} [3n^{2}-l(l +1)]\delta^{2}+\frac{n^{2}}{6}(5n^{2}+1-3l(l+1))\delta^{3}.
  \label{E3}
  \end{equation}
The eigenstate for a given $l=n-s$, with $s=2,3,...,n$, is obtained iteratively using the $\tilde{a}^{(i)}_{n-s}$ operators
\begin{equation}
u_{n,n-s}(x)= N_{n,n-s} (\delta ) a^{(0)}_{n-s}\tilde{a}^{(1)}_{n-s}...\tilde{a}^{(s-2)}_{n-s}\tilde{u}^{(s-1)}_{n,n-s}(x),
\label{wf3}
\end{equation} 
where $a^{(0)}_{n-s}\equiv a_{n-s}$.

This procedure yields the complete set of eigenvalues and eigenstates to the Yukawa problem to order $\delta^{3}$. Concerning eigenstates,
 a systematic calculation to order $\delta^{3}$ requires to expand the solution to this order, which replaces the product of the 
 normalization factor and the exponential in $\tilde{u}^{(s-1)}_{n,n-s}(x)$ with a polynomial of order $\delta^{3}$ with
 $x$-dependent coefficients. In the next section we address this expansion.

 \section{Complete solution to the Yukawa potential to any order in $\delta$}
 \subsection{Energy levels}

The algorithm outlined in the previous section can be applied to any order of the expansion of the Yukawa potential in powers of $\delta$. 
This yields a complete analytical solution to the quantum Yukawa problem. Our calculations show that the shape invariance of the 
potentials is a useful property which, when it holds, make results based on supersymmetry easier to 
obtain. As discussed in the previous section, in general SUSY partner potentials are not shape invariant. In particular, Yukawa potential 
and its SUSY partner are not shape invariant beyond ${\cal O}(\delta^{2})$ but we can still use supersymmetry to completely solve 
the problem to any order in $\delta$. We find that the energy levels depend in general of $n^{2}$ and $l(l+1)$. In the calculation to 
order $\delta^{k}$, the Taylor series for the eigenvalues can be written as
\begin{equation}
\epsilon_{n,l}(\delta )=\sum^{k}_{i=0} \varepsilon_{i}(n^{2},l(l+1))\delta^{i},
\label{Enl}
\end{equation}
and the asymptotic expansion of the function $\epsilon_{n,l}(\delta)$ is obtained for $k\to \infty$.
The coefficients $\varepsilon_{i}(n^{2},l(l+1))$ can be recursively calculated and in the phenomenological analysis below we will 
use this expansion up to order $\delta^{52}$ for the calculation of some observables. However, expressions for the coefficients 
are rather long and we list them here only up to $i=10$. Using the shorthand notation $a=n^{2}$ and $b=l(l+1)$, these coefficients 
are given as
\begin{align}
\varepsilon_{0}(a,b)=&-\frac{1}{a},  \nonumber \\
\varepsilon_{1}(a,b)=& 2,  \nonumber \\
\varepsilon_{2}(a,b)=& -\frac{1}{2}(3a-b),  \nonumber \\
\varepsilon_{3}(a,b)=& \frac{a}{6}(5a-3b+1),  \nonumber\\
\varepsilon_{4}(a,b)=& -\frac{a}{96}(77a^{2}+55a-30ab-15b^{2}-6b), \nonumber \\
\varepsilon_{5}(a,b)=& \frac{a^{2}}{160}(171a^{2}+245a-70ab-45b^{2}-50b+4), \nonumber \\
\varepsilon_{6}(a,b)=&-\frac{a^2 }{2880} \left[4763 a^3-30 a^2 (69 b-386)-7 a \left(135 b^2+420 b-151\right)  \right.\nonumber \\
& \left. -5 b \left(68 b^2+41 b+6\right)\right], \nonumber \\
\varepsilon_{7}(a,b)=& \frac{a^3 }{8064} \left[22763 a^3-77 a^2 (141 b-1100)-21 a \left(195 b^2+1245 b-937\right)  \right.\nonumber \\
& \left.  - 3 \left(721 b^3+1281 b^2+686 b-12\right)\right],\nonumber \\
\varepsilon_{8}(a,b)=& -\frac{a^3}{2580480}\left[ 13283265 a^4-182 a^3 (38034 b-388573)   \right.\nonumber \\
&\left. - 385 a^2 \left(5586 b^2+66312 b-83125\right)  \right.\nonumber \\
 &\left.- 18 a \left(58030 b^3+237265 b^2+303534 b-35598\right)  \right.\nonumber \\
&\left. -105 b \left(2767 b^3 +2228 b^2+580 b+48\right) \right], \nonumber 
\end{align}
\begin{align}
\varepsilon_{9}(a,b)=&\frac{a^4}{3317760}  \left[ 32694383 a^4-30 a^3 (619482 b-7905637) \right. \nonumber \\
&-273 a^2 \left(17910 b^2+359448 b-662119\right) 
 \nonumber \\
   & \quad  -110 a \left(19578 b^3+148791 b^2+381258 b-111470\right)  \nonumber \\
   &\left. -3 \left(337035 b^4+767060 b^3+622580 b^2+211632 b-960\right) \right], \nonumber \\
\varepsilon_{10}(a,b)=&-\frac{a^4}{232243200} \left[ 4546296155 a^5-1020 a^4 (2749521 b-42455893)  \right. \nonumber \\
&-21 a^3 \left(29452110 b^2+961480800 b-2414775527\right)  \nonumber \\
& -130 a^2 \left(1851570 b^3+23985045 b^2+113412222 b-59393407\right) \nonumber \\
& -33 a \left(3815805 b^4+17424260 b^3+31739120 b^2+26224200 b-1200216\right)  \nonumber \\
&\left. -126 b \left(222244 b^4+223865 b^3+82252 b^2+12924 b+720\right) \right].
\label{encoeff}   
\end{align}

\subsection{Eigenstates}
The wave functions are obtained according to Eq.(\ref{wf3}) with the superpotentials $\tilde{W}^{(m)}_{l}(x,\delta )$ calculated to the desired order. 
For a given $n$, the tower of states starts with $u_{n,n-1}$ simply given by the condition $a^{\dagger}_{n-1} u_{n,n-1}=0$, with the solution
\begin{equation}
u_{n,n-1}(x)=N_{n,n-1}(\delta) x^{n}e^{-\frac{x}{n}} e^{-\int w_{n-1}(x,\delta) dx}
\end{equation}
with the complementary superpotential $w_{l}(x,\delta)$ calculated to the desired order $k$. Strictly speaking, this solution is valid to order $k$,
thus we must expand the exponential to obtain
\begin{equation}
u_{n,n-1}(x,\delta,k)=N_{n,n-1}(\delta) x^{n}e^{-\frac{x}{n}} P^{k}_{n,n-1}(x,\delta), 
\end{equation}
where we included explicitly the dependence in $\delta$ and $k$, and $ P^{k}_{n,n-1}(x,\delta)$ is a polynomial of order $k$ in 
$\delta$ with $x$-dependent coefficients, which in turn, are polynomials in $x$.
This structure is preserved when we go to the next levels. In general the solutions to order $k$ have the following structure
\begin{equation}
u_{n,l}(x,\delta,k)=N_{n,l}(\delta) x^{l+1}e^{-\frac{x}{n}} P^{k}_{n,l}(x,\delta). 
\end{equation}
The states must be normalized as
\begin{equation}
\int^{\infty}_{0} dr~ r^{2} |R_{n,l}(r,\delta,k)|^{2}=a^{3}_{0}\int^{\infty}_{0} dx |u_{n,l}(x,\delta,k)|^{2}=1.
\end{equation}
Performing this integration we get
\begin{equation}
\int^{\infty}_{0} |u_{n,l}(x,\delta,k)|^{2}=|N_{n,l}(\delta) |^{2} \left(\frac{n}{2}\right)^{3}\frac{(n+l)! 2n}{(n-l-1)!} \left(\frac{n}{2}\right)^{2l} K_{nl}(\delta)
\end{equation}
where $K_{nl}(\delta)$ is a polynomial of order $\delta^{2k}$. The normalized states are
\begin{equation}
u_{n,l}(x,\delta,k)=\eta_{nl} \sqrt{\left(\frac{2}{n a_{0}}\right)^{3}\frac{(n-l-1)!}{(n+l)! 2n} }  \left(\frac{2}{n}\right)^{l} \frac{1}{\sqrt{K_{nl}(\delta)}}   
x^{l+1}e^{-\frac{x}{n}} P^{k}_{n,l}(x,\delta), 
\end{equation}
where $\eta_{nl}$ are phases. A systematic calculation of the eigenstates to order $\delta^{k}$ requires to expand also the normalization
factor in the above expressions to this order which yields
\begin{equation}
u_{n,l}(x,\delta,k)=\sqrt{\left(\frac{2}{n a_{0}}\right)^{3}\frac{(n-l-1)!}{(n+l)! 2n} }  \left(\frac{2}{n}\right)^{l}   
x^{l+1}e^{-\frac{x}{n}} M^{k}_{n,l}(x,\delta),
\end{equation}
where $M^{k}_{n,l}(x,\delta)$ are new polynomials of order $k$ in $\delta$ which incorporate additional $\delta$ contributions from the 
normalization factors. Choosing the phases as $\eta_{nl}=(-1)^{n-l-1}$, in the $\delta \to 0$ limit these polynomials reduce to
\begin{equation}
M^{k}_{n,l}(x,0)= L^{2l+1}_{n-l-1}\left(\frac{2x}{n}\right),
\end{equation}
where $L^{2l+1}_{n-l-1}(\rho)$ stands for the Laguerre associated polynomials. With the above choice of phases we recover the well known 
solutions to the Coulomb problem for $\delta= 0$.

Summarizing, in terms of $\rho=2x/n=2r/na_{0}$, the eigenstates of the Yukawa potential to order $\delta^{k}$ are given by 
\begin{equation}
\psi_{nlm}(\mathbf{r},\delta,k)=\sqrt{\left(\frac{2}{n a_{0}}\right)^{3}\frac{(n-l-1)!}{(n+l)! 2n} }   \rho^{l}e^{-\frac{\rho}{2}} 
N^{2l+1}_{n-l-1}(\rho,\delta,k) Y^{m}_{l}(\theta,\phi),
\label{eigenstates}
\end{equation} 
 where
 \begin{equation}
 N^{2l+1}_{n-l-1}(\rho,\delta,k)=M^{k}_{n,l}(\frac{n\rho}{2},\delta).
 \end{equation}
 The explicit form of these polynomials are easily calculated to the desired order $k$. For future reference,
  in the Appendix we list these polynomials to order $\delta^{5}$ for the lowest lying levels, $n=1,2,3,4$.
\section{Discussion}
\subsection{Comparison with existing analytical results}
There are many calculations with partial results of the energy levels of the Yukawa potentials in the literature but most of them focus on 
the ground state or the first few levels. Also, only a few of them attempt a systematic expansion and yield results to higher order with 
which we can compare our results. We skip comparison with calculations based on variational principles for this reason. Analytical results 
to order $\delta^{2}$ for the Yukawa potential were obtained in \cite{PhysRevA.13.532} using a variation of perturbation theory named 
analytic perturbation theory. Our results to order $\delta^{2}$ reproduce results in that work. Using first order conventional perturbation 
theory with the Coulomb potential as the unperturbed system and approximate calculation of the involved matrix elements, the energy 
levels of the Yukawa potential were calculated in \cite{PhysRev.134.A1235} up to order $\delta^{5}$. Our results agree with results in that work   
up to order $\delta^{3}$. The calculation of the ground state energy has been addressed by many authors. The most elaborated 
perturbative calculation for $n=1$ has been done  in \cite{Eletsky:1981fm} \cite{Vainberg:1981} using Logarithmic Perturbation Theory. 
Our results for the ground state agree with these calculations. A more comprehensive calculation of levels was done in \cite{Gonul:2006} 
using related techniques, where calculations for $n=1,2,3$ levels are done up to order $\delta^{4}$. 
Our calculation also reproduce these results.

\begin{figure}[h]
\begin{center}
\includegraphics[scale=0.55]{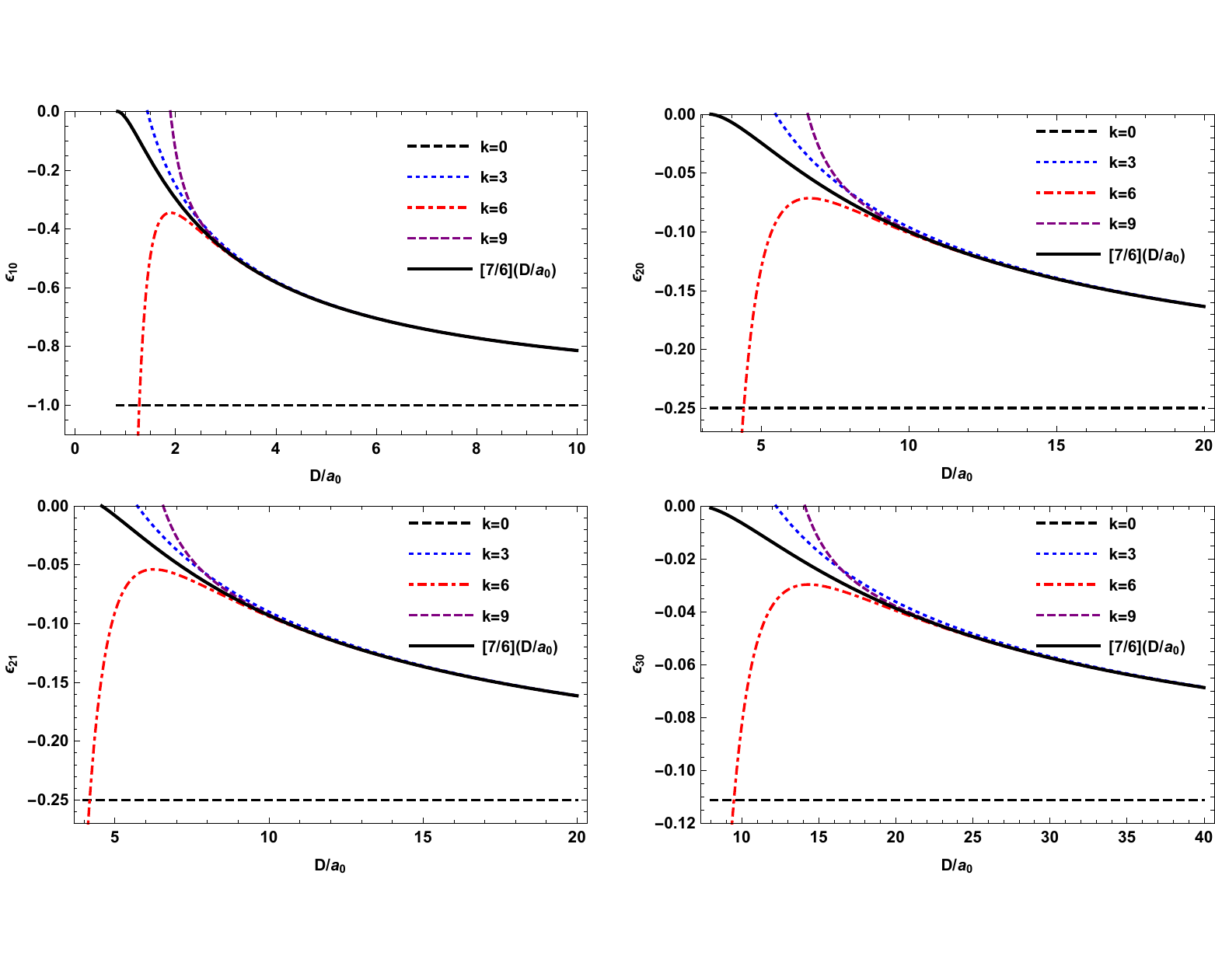}
\end{center}
\caption{Lowest energy levels of the Yukawa potential (in Rydberg units) as functions of the screening length $D$, calculated at different 
order $k$ in the perturbative expansion, and the $[(N+1)/N]$ Pad\'e approximant for $N=6$.}
\label{epsnlk}
\end{figure}

\subsection{Analysis of the energy levels}
The coefficients $\varepsilon_{i}(n^{2},l(l+1))$ in Eqs. (\ref{encoeff}) grow with $i$.  This raises the concern on the convergence radii 
of the series for the energy levels in Eq.(\ref{Enl}). The convergence of Taylor series arising in perturbative calculations has been studied in 
\cite{PhysRev.184.1231}\cite{Bender:1973rz}\cite{ZinnJustin:1980uk}\cite{PhysRevD.23.2916} \cite{Okopinska:1987hp} and several 
methods have been devised to construct analytical extensions beyond the corresponding convergence radii \cite{Arteca:1990xe}. 
We will work here with the Pad\'{e} approximants technique \cite{osti_4454325}. For the energy level expansion in Eq.(\ref{Enl}) to 
order $k=M+N$ it is always possible to construct a rational function
\begin{equation}
[M/N](\delta)=\frac{P_{M}(\delta)}{Q_{N}(\delta)}
\end{equation} 
with $P_{M}(\delta)$ and $Q_{N}(\delta)$ polynomials of order $M$ and $N$ respectively. The Taylor expansion of the $[M/N]$ Pad\'{e} 
approximant coincides with the  expansion of $\epsilon_{nl}(\delta)$ to order $k=M+N$. The coefficients of the $P_{M}$ and $Q_{N}$ 
polynomials are fixed by the coefficients $\varepsilon_{i}(n^{2},l(l+1))$ and these polynomials are unique. The value of the series is bounded 
from above and below by the values of the $[(N+1)/N]$ and $[N/N]$ approximants  \cite{Arteca:1990xe}. As we go to high $N$, these 
approximants converge \cite{Napsuciale:2020ehf} and we can estimate the uncertainty in the calculation of the levels as the 
difference in their values for a given $\delta$.  In Fig. \ref{epsnlk} we plot the energy levels for the lower states as functions of the 
screening length $D$ (in units of the Bohr radius) 
calculated to order $k=0,3,6,9$, and the $[(N+1)/N]$ Pad\'{e} approximant for $N=6$. We use the {\it Mathematica} package for the calculation 
of the Pad\'e approximants in this paper. We can see in these plots that the Taylor series yields a good description everywhere except 
close to the critical screening, while the $[(N+1)/N]$ and $[N/N]$ approximants, enlarge the convergence radii up to the critical region. 
The precision in the calculation of the ground state energy for $D/a_{0}=1$ and $N=6$ is 
$[7/6](1)-[6/6](1)=9.8\times 10^{-5}$ . In general, 
whenever we are far from the critical region, results to order $\delta^{3}$ yield a good description of the system. The approximation 
at this order fails as we approach the critical region and results for the Taylor series at higher orders do not improve this description. Here, 
the use of the Pad\'{e} approximants is necessary and a more precise calculation is reached as we increase the $N$ of the approximant, 
which requires a calculation to higher order of the Taylor series. For $N=6$ we reach the precision for the energy levels in the critical region 
obtained in the numerical solution of the Yukawa potential reported in  \cite{PhysRevA.1.1577}, and our calculations reproduce 
the  complete set of numerical results for $n=1,..,9$ given there. 
The precision in the calculation of the energy levels can be further improved taking higher $N$ approximants, e.g., for the ground state 
and $N=10$, which requires to calculate the Taylor series to order $k=21$, we find $[11/10](1)-[10/10](1)=9\times 10^{-8}$ and this 
uncertainty in the calculation is considerably lower in the low $\delta$ (large screening) region.
\begin{figure}[h]
\begin{center}
\includegraphics[scale=0.55]{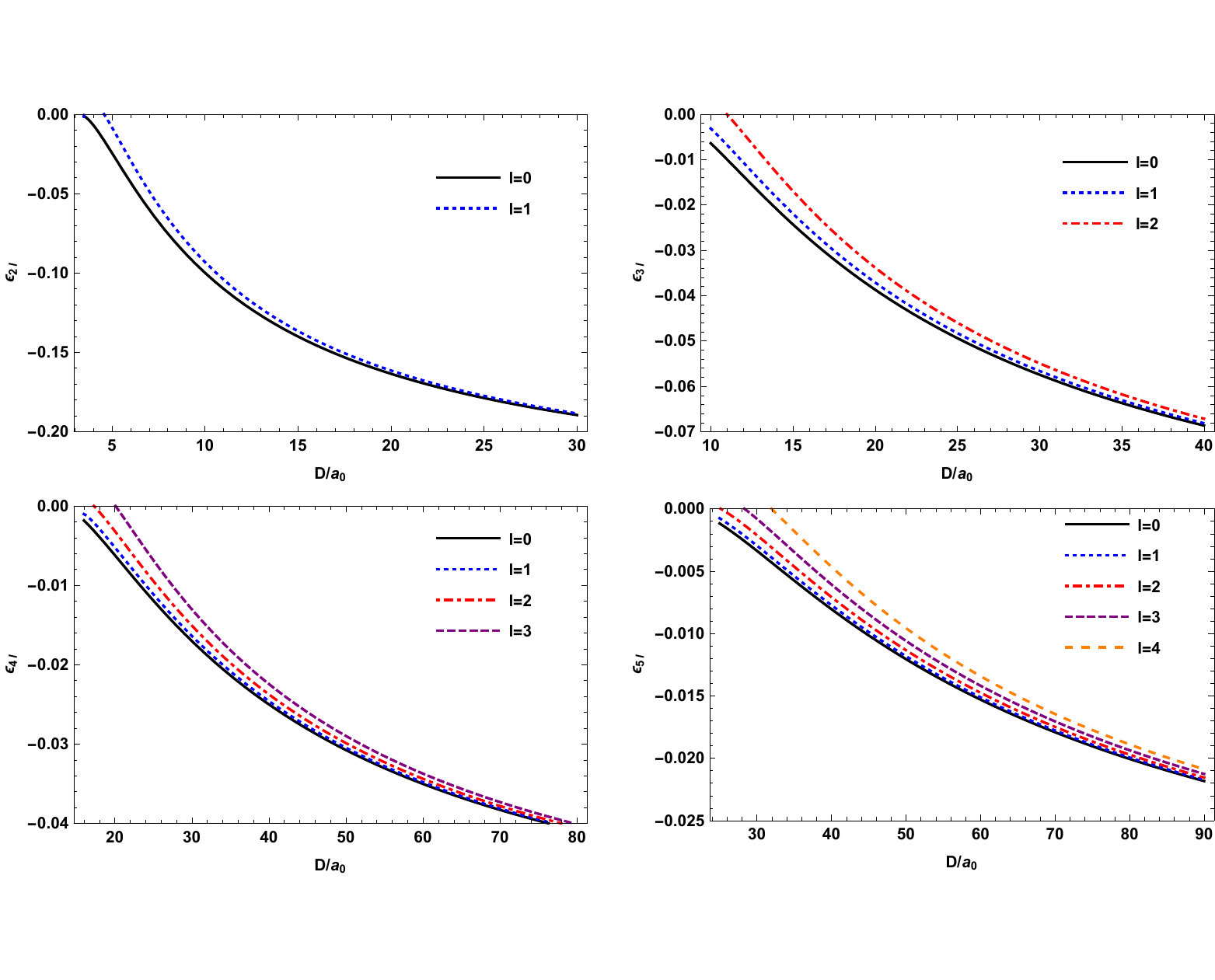}
\end{center}
\caption{Lowest energy levels of the Yukawa potential (in Rydberg units), $\epsilon_{nl}$, as functions of the screening length $D$ (in Bohr
radius units) calculated using the $[(N+1)/N]$ Pad\'e approximant with $N=6$.}
\label{epsnl}
\end{figure}

\begin{figure}[h]
\begin{center}
\includegraphics[scale=0.55]{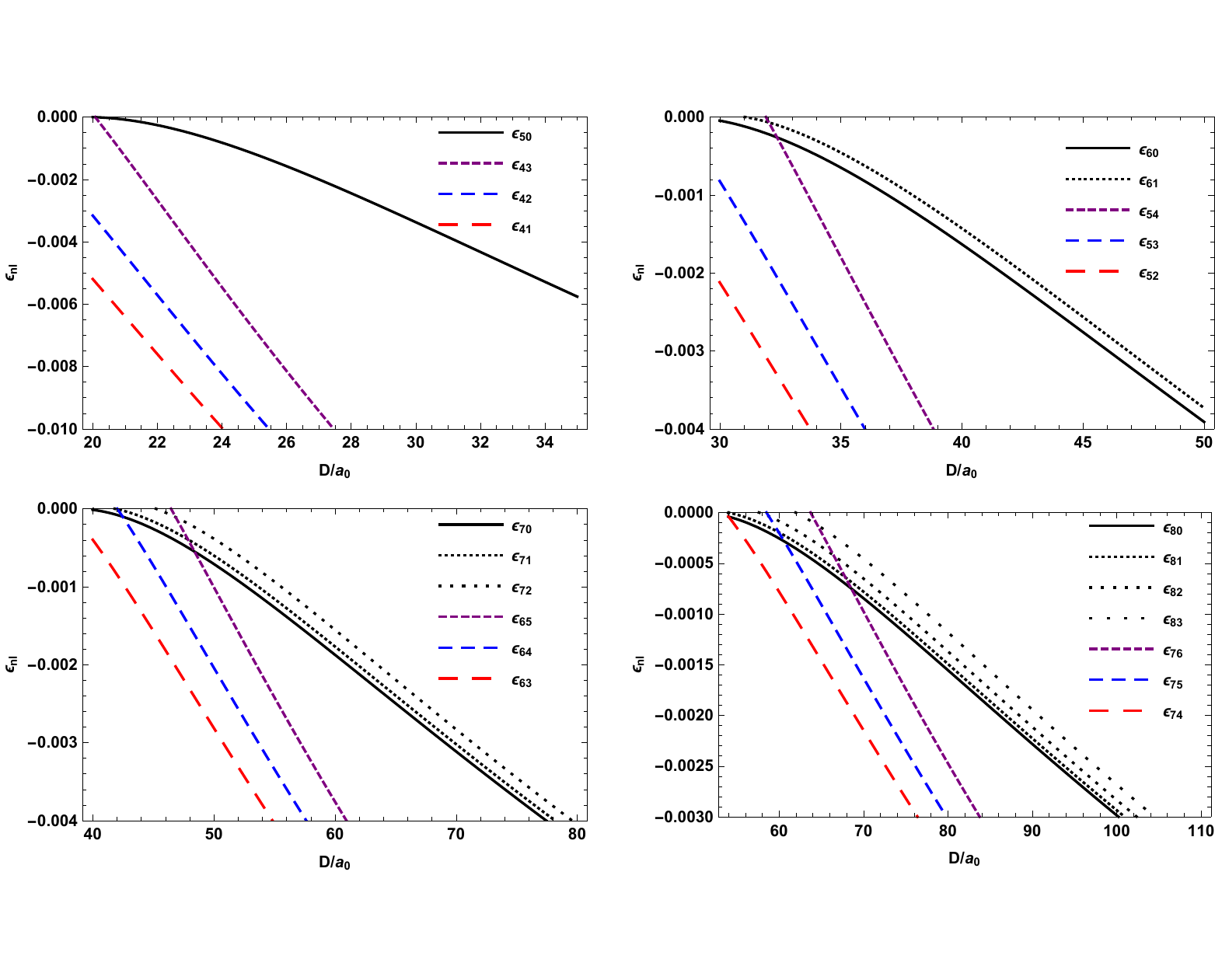}
\end{center}
\caption{Crossing of  energy levels of the Yukawa potential. Curves correspond to energy levels calculated using the $[(N+1)/N]$ 
Pad\'e approximant with $N=10$.}
\label{Crossing}
\end{figure}

In general we find that, for a given $n$, the higher $l$ states have higher energy. This is shown in Fig. \ref{epsnl}, where we use the 
$[7/6](\delta)$ Pad\'{e} approximant in the calculation of the energy levels and plot them as functions of the screening length measured 
in units of the Bohr radius. Also, as we go to high values of $n$, the gap between the energies of the $l=0$ and $l=n-1$ increases and 
eventually there is a cross-over of the energy levels, i.e.,  $\epsilon_{n,l}>\epsilon_{n+1,l^{\prime}}$. 
We find that the lowest $n$ for which this phenomena occurs is $n=4$. For low $n$ the crossing of levels takes place near the critical regions and 
to clearly see it is necessary to go to higher orders in the perturbative expansion.  In Fig. \ref{Crossing} we show results of the calculation of 
the energy levels for $n=4,5,6,7,8$ using the $[(N+1)/N]$ Pad\'e approximant with $N=10$ which requires a calculation of the series to order $k=21$. 

\subsection{Critical screening lengths}
The critical screening lengths $\delta_{nl}$,  defined as the values of $\delta$ at which the levels $\epsilon_{n,l}$ go to the continuum, are 
very important for practical applications. They are in the large $\delta$ region and we must use the $[(N+1)/N]$ and
$[N/N]$ Pad\'{e} approximants in their calculation. For a given level $\epsilon_{nl}$, we find the values of $\delta\equiv \delta^{N+1}_{nl}$ and 
$\delta\equiv \delta^{N}_{nl}$ at which $[(N+1)/N](\delta^{N+1}_{nl})=0$ and $[N/N](\delta^{N}_{nl})=0$, respectively. For large enough values of $N$, these 
values of $\delta$ coincide up to a given figure and the actual value of $\delta_{nl}$ lies between $\delta^{N+1}_{nl}$ and $\delta^{N}_{nl}$ 
which yields the uncertainty in the calculation. 

There are two important considerations in this procedure. First, each level has its own Taylor series and consequently its own set of 
Pad\'e approximants in the calculation to a given order $k=2N+1$. In principle we can use a common value of $N$ in the calculation
of all $\epsilon_{nl}$, but this yields different uncertainties for the different levels. Second,  the Pad\'{e} approximants are rational functions 
of two polynomials, thus they have poles, some of them on the real axis. It sometimes happens that, for a given $N$, some of these real poles are   
in the physical region and close to the critical values. The reliable determination of critical screening lengths, requires to use approximants 
with no poles in the physical regions. With these considerations we obtain the critical screening lengths listed in Tabel \ref{csl}. The crossing 
of energy levels can also be  seen in this table, where we notice that $\delta_{43}<\delta_{50}$. Whenever $\delta_{n,l}<\delta_{n+1,l^{\prime}}$ we have this phenomena and as we go to higher $n$ more and more $n+1$ levels are crossed by the $n$ levels.
\begin{table}
\begin{center}
\caption{Critical screening lengths, $\delta_{nl}$, for $n=1,...,9$, calculated with the $[(N+1)/N]$ and $[N/N]$ Pad\'e approximants. For comparison
we also show results from the numerical solutions given in Ref.\cite{PhysRevA.1.1577}(see also \cite{Diaz_1991}). }
 \label{csl}
\bigskip 
\begin{tabular}{||c|c|c|c|c||ccccc}
\cline{1-5}\cline{6-10}
$n$ & $l$ & $N$ & $\delta _{nl}$ & Numeric & $n$ & \multicolumn{1}{|c}{$l$}
& \multicolumn{1}{|c}{$N$} & \multicolumn{1}{|c}{$\delta _{nl}$} & 
\multicolumn{1}{|c||}{Numeric} \\ \cline{1-5}\cline{6-10}
$1$ & $0$ & $21$ & \multicolumn{1}{|l|}{$1.1906124207\left( 2\right) $} & 
\multicolumn{1}{|l||}{$1.1906$} &  & \multicolumn{1}{|c}{$0$} & 
\multicolumn{1}{|c}{$23$} & \multicolumn{1}{|l|}{$0.0198221\left( 1\right) $}
& \multicolumn{1}{|l||}{$0.0198$} \\ \cline{1-5}
$2$ & $0$ & $24$ & \multicolumn{1}{|l|}{$0.3102092834\left( 2\right) $} & 
\multicolumn{1}{|l||}{$0.3103$} &  & \multicolumn{1}{|c}{$1$} & 
\multicolumn{1}{|c}{$24$} & \multicolumn{1}{|l|}{$0.01862667\left( 2\right) $%
} & \multicolumn{1}{|l||}{$0.0186$} \\ 
& $1$ & $21$ & \multicolumn{1}{|l|}{$0.220118\left( 1\right) $} & 
\multicolumn{1}{|l||}{$0.2202$} &  & \multicolumn{1}{|c}{$2$} & 
\multicolumn{1}{|c}{$23$} & \multicolumn{1}{|l|}{$0.01738685\left( 4\right) $%
} & \multicolumn{1}{|l||}{$0.0174$} \\ \cline{1-5}\cline{5-5}
& $0$ & $21$ & \multicolumn{1}{|l|}{$0.139450295\left( 1\right) $} & 
\multicolumn{1}{|l||}{$0.1395$} & $8$ & \multicolumn{1}{|c}{$3$} & 
\multicolumn{1}{|c}{$25$} & \multicolumn{1}{|l|}{$0.01615594\left( 4\right) $%
} & \multicolumn{1}{|l||}{$0.0162$} \\ 
$3$ & $1$ & $26$ & \multicolumn{1}{|l|}{$0.11265\left( 1\right) $} & 
\multicolumn{1}{|l||}{$0.1127$} &  & \multicolumn{1}{|c}{$4$} & 
\multicolumn{1}{|c}{$24$} & \multicolumn{1}{|l|}{$0.01498071\left( 1\right) $%
} & \multicolumn{1}{|l||}{$0.0150$} \\ 
& $2$ & $22$ & \multicolumn{1}{|l|}{$0.0913384\left( 2\right) $} & 
\multicolumn{1}{|l||}{$0.0914$} &  & \multicolumn{1}{|c}{$5$} & 
\multicolumn{1}{|c}{$23$} & \multicolumn{1}{|l|}{$0.01388348\left( 1\right) $%
} & \multicolumn{1}{|l||}{$0.0139$} \\ \cline{1-5}\cline{1-1}\cline{4-4}
& \multicolumn{1}{|c|}{$0$} & $24$ & \multicolumn{1}{|l|}{$%
0.0788281106\left( 1\right) $} & \multicolumn{1}{|l||}{$0.0788$} &  & 
\multicolumn{1}{|c}{$6$} & \multicolumn{1}{|c}{$21$} & \multicolumn{1}{|l|}{$%
0.012871446\left( 2\right) $} & \multicolumn{1}{|l||}{$0.0129$} \\ 
$4$ & \multicolumn{1}{|c|}{$1$} & $21$ & \multicolumn{1}{|l|}{$%
0.067827\left( 1\right) $} & \multicolumn{1}{|l||}{$0.0679$} &  & 
\multicolumn{1}{|c}{$7$} & \multicolumn{1}{|c}{$21$} & \multicolumn{1}{|l}{$%
0.011944528\left( 1\right) $} & \multicolumn{1}{|l||}{$0.0119$} \\ 
\cline{6-10}
& \multicolumn{1}{|c|}{$2$} & $22$ & \multicolumn{1}{|l|}{$0.058099\left(
1\right) $} & \multicolumn{1}{|l||}{$0.0581$} &  & \multicolumn{1}{|c}{$0$}
& \multicolumn{1}{|c}{$25$} & \multicolumn{1}{|l|}{$0.0156708\left( 1\right) 
$} & \multicolumn{1}{|l||}{$0.0157$} \\ 
& \multicolumn{1}{|c|}{$3$} & \multicolumn{1}{|c|}{$23$} & 
\multicolumn{1}{|l|}{$0.049830665\left( 4\right) $} & \multicolumn{1}{|l||}{$%
0.0498$} &  & \multicolumn{1}{|c}{$1$} & \multicolumn{1}{|c}{$25$} & 
\multicolumn{1}{|l|}{$0.0148561\left( 1\right) $} & \multicolumn{1}{|l||}{$%
0.0149$} \\ \cline{1-5}\cline{1-5}
& \multicolumn{1}{|c|}{$0$} & \multicolumn{1}{|c|}{$25$} & 
\multicolumn{1}{|l|}{$0.0505831707\left( 2\right) $} & \multicolumn{1}{|l||}{%
$0.0506$} &  & \multicolumn{1}{|c}{$2$} & \multicolumn{1}{|c}{$24$} & 
\multicolumn{1}{|l|}{$0.01399716\left( 1\right) $} & \multicolumn{1}{|l||}{$%
0.0140$} \\ 
& \multicolumn{1}{|c|}{$1$} & \multicolumn{1}{|c|}{$24$} & 
\multicolumn{1}{|l|}{$0.045155\left( 1\right) $} & \multicolumn{1}{|l||}{$%
0.0452$} &  & \multicolumn{1}{|c}{$3$} & \multicolumn{1}{|c}{$23$} & 
\multicolumn{1}{|l|}{$0.01312892\left( 2\right) $} & \multicolumn{1}{|l||}{$%
0.0131$} \\ 
$5$ & \multicolumn{1}{|c|}{$2$} & \multicolumn{1}{|c|}{$23$} & 
\multicolumn{1}{|l|}{$0.0400197\left( 1\right) $} & \multicolumn{1}{|l||}{$%
0.0400$} & $9$ & \multicolumn{1}{|c}{$4$} & \multicolumn{1}{|c}{$22$} & 
\multicolumn{1}{|l|}{$0.01228586\left( 1\right) $} & \multicolumn{1}{|l||}{$%
0.0123$} \\ 
& \multicolumn{1}{|c|}{$3$} & \multicolumn{1}{|c|}{$21$} & 
\multicolumn{1}{|l|}{$0.0353883\left( 1\right) $} & \multicolumn{1}{|l||}{$%
0.0354$} &  & \multicolumn{1}{|c}{$5$} & \multicolumn{1}{|c}{$23$} & 
\multicolumn{1}{|l|}{$0.011485698\left( 1\right) $} & \multicolumn{1}{|l||}{$%
0.0115$} \\ 
& \multicolumn{1}{|c|}{$4$} & \multicolumn{1}{|c|}{$22$} & 
\multicolumn{1}{|l|}{$0.031343456\left( 1\right) $} & \multicolumn{1}{|l||}{$%
0.0313$} &  & \multicolumn{1}{|c}{$6$} & \multicolumn{1}{|c}{$22$} & 
\multicolumn{1}{|l}{$0.010736127\left( 1\right) $} & \multicolumn{1}{|l||}{$%
0.0107$} \\ \cline{1-5}\cline{3-3}\cline{5-5}
& \multicolumn{1}{|c|}{$0$} & \multicolumn{1}{|c|}{$26$} & 
\multicolumn{1}{|l|}{$0.035183478\left( 1\right) $} & \multicolumn{1}{|l||}{$%
0.0352$} &  & \multicolumn{1}{|c}{$7$} & \multicolumn{1}{|c}{$21$} & 
\multicolumn{1}{|l}{$0.0100397512\left( 1\right) $} & \multicolumn{1}{|l||}{$%
0.0100$} \\ 
& \multicolumn{1}{|c|}{$1$} & \multicolumn{1}{|c|}{$26$} & 
\multicolumn{1}{|l|}{$0.0321562\left( 2\right) $} & \multicolumn{1}{|l||}{$%
0.0322$} &  & \multicolumn{1}{|c}{$8$} & \multicolumn{1}{|c}{$23$} & 
\multicolumn{1}{|l}{$0.0093959992\left( 1\right) $} & \multicolumn{1}{|l||}{$%
0.0094$} \\ \cline{6-10}
$6$ & \multicolumn{1}{|c|}{$2$} & \multicolumn{1}{|c|}{$23$} & 
\multicolumn{1}{|l|}{$0.0291623\left( 2\right) $} & \multicolumn{1}{|l||}{$%
0.0292$} &  &  &  & \multicolumn{1}{l}{} & \multicolumn{1}{l}{} \\ 
& \multicolumn{1}{|c|}{$3$} & \multicolumn{1}{|c|}{$24$} & 
\multicolumn{1}{|l|}{$0.02635015\left( 2\right) $} & \multicolumn{1}{|l||}{$%
0.0264$} &  &  &  & \multicolumn{1}{l}{} & \multicolumn{1}{l}{} \\ 
& \multicolumn{1}{|c|}{$4$} & \multicolumn{1}{|c|}{$22$} & 
\multicolumn{1}{|l|}{$0.02379897\left( 2\right) $} & \multicolumn{1}{|l||}{$%
0.0238$} &  &  &  & \multicolumn{1}{l}{} & \multicolumn{1}{l}{} \\ 
& $5$ & \multicolumn{1}{|c|}{$22$} & \multicolumn{1}{|l|}{$0.021524523\left(
3\right) $} & \multicolumn{1}{|l||}{$0.0215$} &  &  &  & \multicolumn{1}{l}{}
& \multicolumn{1}{l}{} \\ \cline{1-3}\cline{3-5}\cline{4-5}
\multicolumn{1}{||l|}{} & \multicolumn{1}{|l|}{$0$} & \multicolumn{1}{|l|}{$%
21$} & \multicolumn{1}{|l|}{$0.02586938\left( 2\right) $} & 
\multicolumn{1}{|l||}{$0.0260$} &  &  &  &  &  \\ 
\multicolumn{1}{||l|}{} & \multicolumn{1}{|l|}{$1$} & \multicolumn{1}{|l|}{$%
22$} & \multicolumn{1}{|l|}{$0.024026435\left( 1\right) $} & 
\multicolumn{1}{|l||}{$0.0240$} &  &  &  &  &  \\ 
\multicolumn{1}{||l|}{} & \multicolumn{1}{|l|}{$2$} & \multicolumn{1}{|l|}{$%
26$} & \multicolumn{1}{|l|}{$0.0221591\left( 1\right) $} & 
\multicolumn{1}{|l||}{$0.0222$} &  &  &  &  &  \\ 
\multicolumn{1}{||l|}{$7$} & \multicolumn{1}{|l|}{$3$} & 
\multicolumn{1}{|l|}{$22$} & \multicolumn{1}{|l|}{$0.02034124\left( 1\right) 
$} & \multicolumn{1}{|l||}{$0.0203$} &  &  &  &  &  \\ 
\multicolumn{1}{||l|}{} & \multicolumn{1}{|l|}{$4$} & \multicolumn{1}{|l|}{$%
23$} & \multicolumn{1}{|l|}{$0.01864606\left( 1\right) $} & 
\multicolumn{1}{|l||}{$0.0186$} &  &  &  &  &  \\ 
\multicolumn{1}{||l|}{} & \multicolumn{1}{|l|}{$5$} & \multicolumn{1}{|l|}{$%
22$} & \multicolumn{1}{|l|}{$0.017095093\left( 1\right) $} & 
\multicolumn{1}{|l||}{$0.0171$} &  &  &  &  &  \\ 
\multicolumn{1}{||l|}{} & \multicolumn{1}{|l|}{$6$} & \multicolumn{1}{|l|}{$%
22$} & \multicolumn{1}{|l|}{$0.015691075\left( 2\right) $} & 
\multicolumn{1}{|l||}{$0.0157$} &  &  &  &  &  \\ 
\cline{1-3}\cline{3-5}\cline{4-5}
\end{tabular}
\end{center}
\end{table}
\subsection{Wavefunctions}
\begin{figure}[h]
\begin{center}
\includegraphics[scale=0.6]{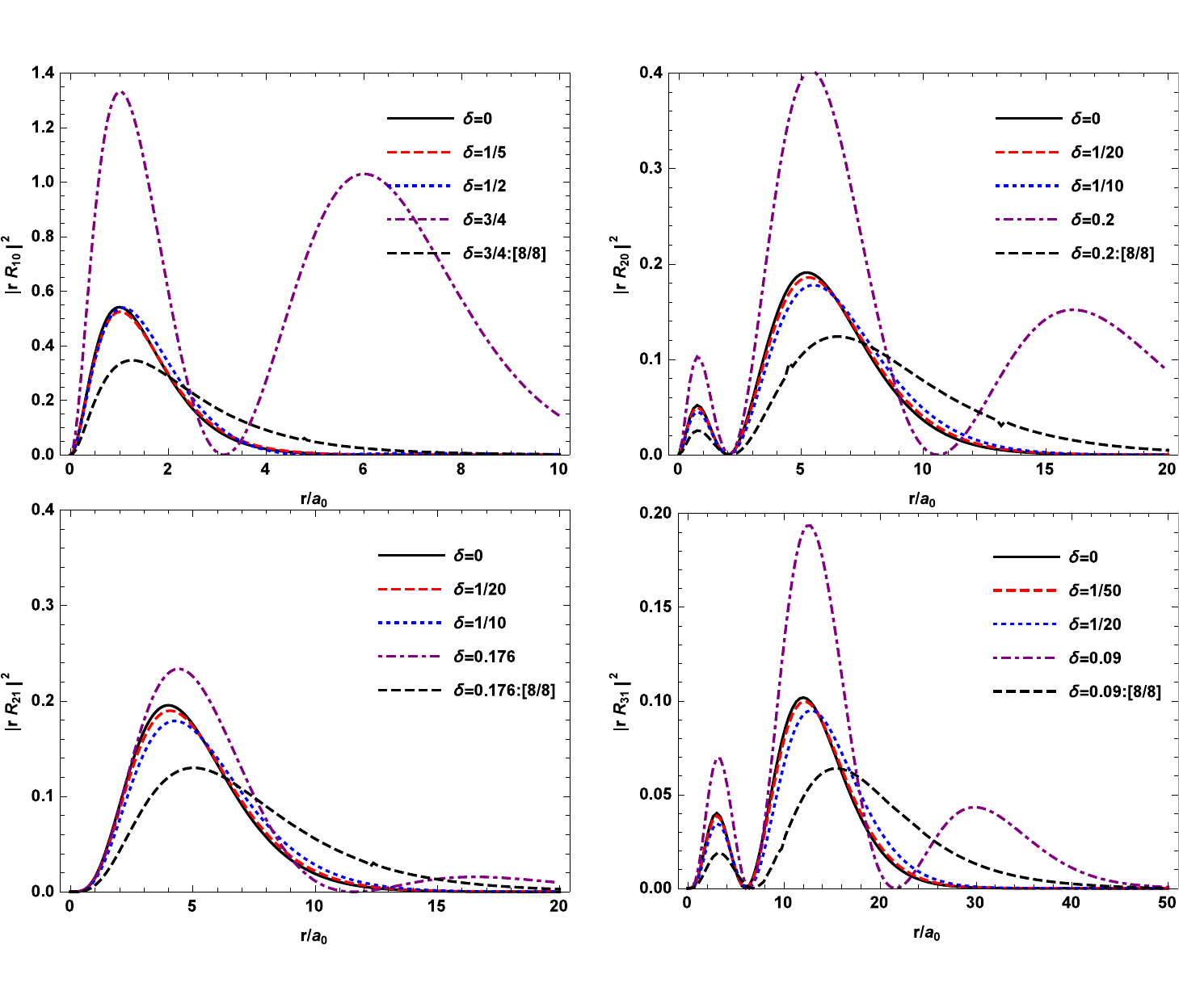}
\end{center}
\caption{Radial probabilities for the lowest states as functions of $\delta$ calculated to order $k=5$ in the perturbative expansion.  Curves 
labelled as $[N/N]$ correspond to the Pad\'e approximant with $N=8$ and $\delta$ close to the critical region.}
\label{Probnl}
\end{figure}

The wave functions for the Yukawa potential calculated to a given order $k$ in the expansion of the potential in powers of $\delta$ are given 
in Eq.(\ref{eigenstates}).  Notice that we incorporated the $\delta$ dependent normalization factor in the $N^{2l+1}_{n-l-1}(x,\delta,k)$ 
polynomials and calculate observables to ${\cal O}(\delta^{k}$), thus these states are normalized as
\begin{equation}
\int d^{3}r |\psi_{nlm}(\mathbf{r},\delta,k)|^{2}=1+{\cal O}(\delta^{k+1}).
\end{equation}
This requires that the $N^{2l+1}_{n-l-1}(x,\delta,k)$ polynomials, even if they are $\delta$-dependent, satisfy the following 
normalization relations 
\begin{equation}
\int^{\infty}_{0}d\rho \,  \rho^{g+1}e^{-\rho} (N^{g}_{r}(\rho,\delta,k ))^{2}
= \frac{(2r+g+1)  (r+g)!}{r!}  + {\cal O}(\delta^{k+1}).
\end{equation}
We expect that eigenstates in Eq.(\ref{eigenstates}) yield a good description of the system whenever we are not close to the critical 
screening region, where terms of ${\cal O}(\delta^{k+1})$ become important and we loose 
the appropriate normalization. In Fig. \ref{Probnl} we show the radial probabilities for the lowest eigenstates, where we can see that proper 
normalization is lost for $\delta\approx 3\delta_{nl}/4$. Nevertheless, similarly to the case of the energy levels, we can construct the 
Pad\'{e} approximants for the solutions to enlarge their convergence radii. 
In this case, the coefficients of the polynomials in $\delta$ entering the Pad\'{e} approximant are in turn quotients of
polynomials in $x$ which makes the calculation of the Pad\'{e} approximants more resource consuming and with conventional 
computing capabilities we cannot go very high in the order of the approximants. However, we can go high enough ($N=8$) to reach the 
beginning of the critical screening region ($\delta \approx 3 \delta_{nl}/4$) as shown in Fig. \ref{Probnl}. As it is clear in these plots, 
a delocalization phenomenon starts taking place for these values of $\delta$. 
It is interesting to notice that even for not so small $\delta$ the radial probabilities are quite similar to those of the Coulomb potential, 
screening becoming important only near the critical region. This is in contrast with screening effects in the energy levels which depart 
substantially for the Coulomb values even for small values of $\delta$. 

In cold dark matter phenomenology, dark matter is non-relativistic and the exchange of massive dark gauge fields in the non-relativistic 
domain produces a Yukawa potential with $\alpha_{D}=g^{2}_{D}/4\pi$ and $\delta=a_{0}/D=2M_{GB}/m\alpha_{g}$, where 
$g_{D}$ stands for the dark matter-dark matter-gauge boson coupling, $M_{GB}$ denotes the dark gauge boson mass and $m$ is 
the dark matter mass (the reduced mass is $\mu=m/2$). Whenever  $\delta=2M_{GB}/m\alpha_{g} < \delta_{10}$, where 
$\delta_{10}= 1.1906124207\left( 2\right)$ is the ground state critical screening length,  there will be non-relativistic dark matter bound 
states (darkonium). This requires a dark gauge boson with a mass $M_{GB}< m\alpha_{g} \delta_{10}/2$, which considering 
perturbative interactions yields a light gauge boson compared with the dark matter mass.
  
\begin{figure}[h]
\begin{center}
\includegraphics[scale=0.65]{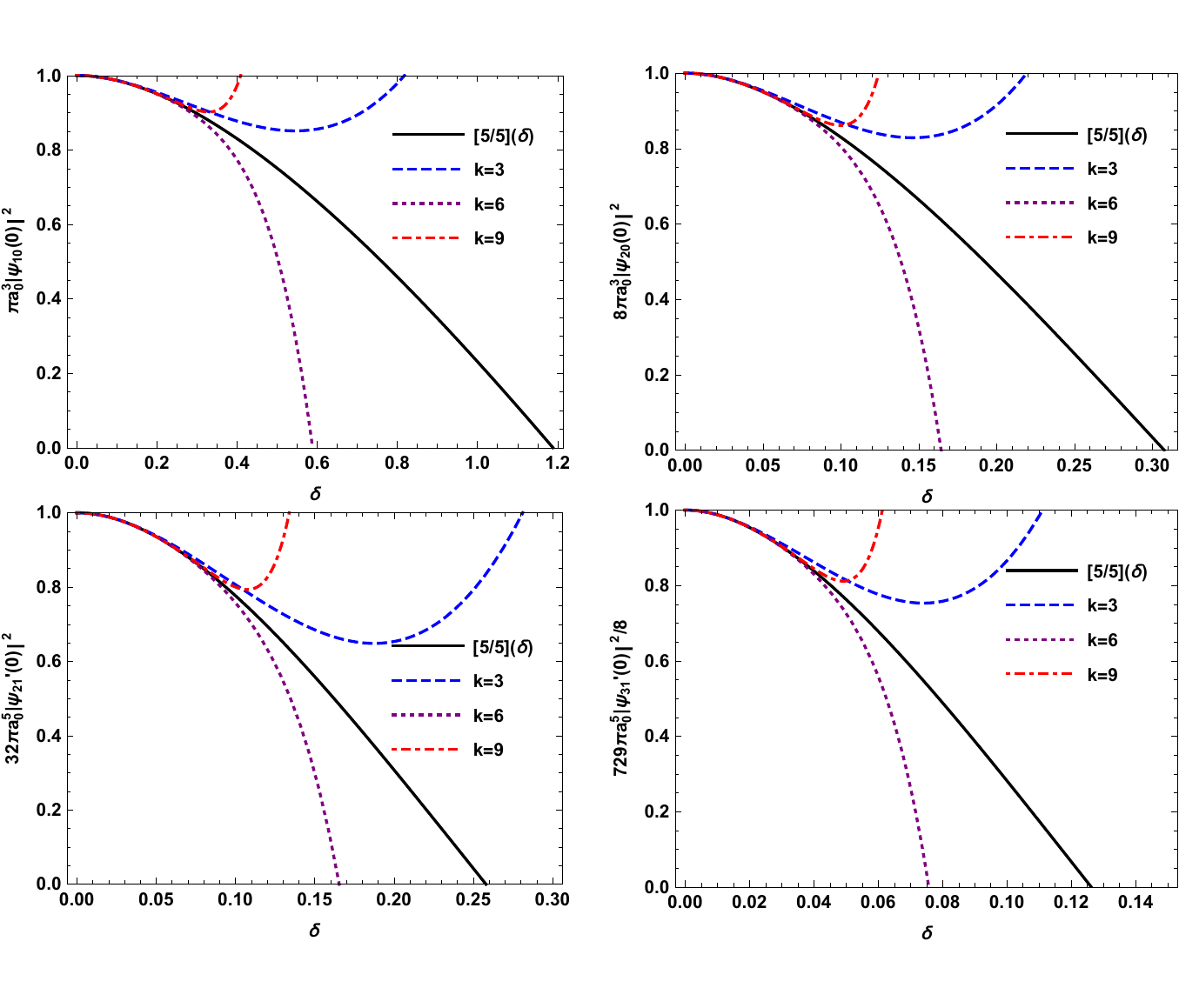}
\end{center}
\caption{Squared wave functions (for $l=0$) and radial derivatives (for $l=1$) evaluated at the origin  for the lowest lying states as 
functions of $\delta$ calculated to order $k=10$ in the perturbative expansion.  Curves 
labelled as $[5/5](\delta)$ correspond to the Pad\'e approximant calculation with $N=5$.}
\label{WF2nl}
\end{figure}

 Darkonium phenomenology  for $s$-wave bound states involves the value of the wave function at the origin 
 $\psi_{n00}(0)=R_{n0}(0)/\sqrt{4\pi} $, while for $p$-wave, observables depend on the value of the radial derivative 
 of the wave function evaluated at the the origin $\psi^{\prime}_{n10}(0)=\sqrt{3/4\pi}R^{\prime}_{n1}(0)$ \cite{Guberina:1980dc}. 
 A calculation to order $\delta^{10}$ for the lower states yields
\begin{align}
|\psi_{100}(0)|^{2}&=\frac{1}{\pi a^{3}_{0}}\left(1-\frac{3}{2}\delta^{2}+\frac{11}{6}\delta^{3}-\frac{341}{96}\delta^{4}+\frac{1427}{160}\delta^{5} 
-\frac{36653}{1440}\delta^{6}+\frac{319447}{4032}\delta^{7} \right. \nonumber \\
&\left.-\frac{169970813}{645120}\delta^{8} +\frac{63908537}{69120}\delta^{9}-\frac{2434733481}{716800}\delta^{10} \right), \\
|\psi_{200}(0)|^{2}&=\frac{1}{8\pi a^{3}_{0}}\left(1-24 \delta^{2}+\frac{328}{3}\delta^{3}-\frac{2068}{3}\delta^{4}+\frac{27784}{5}\delta^{5} 
-\frac{2277964}{45}\delta^{6} \right. \nonumber \\
&\left. +\frac{31663864}{63}\delta^{7}-\frac{3345668933}{630}\delta^{8} +\frac{7950329914}{135}\delta^{9}-\frac{50941652654}{75}\delta^{10} \right), 
\end{align} 
\begin{align}
|\psi^{\prime}_{210}(0)|^{2}&=
\frac{1}{32\pi a^{5}_{0}}\left(1-30 \delta^{2}+\frac{320}{3}\delta^{3}-\frac{1550}{3}\delta^{4}+\frac{20824}{5}\delta^{5} 
-\frac{192536}{5}\delta^{6}  \right. \nonumber \\
&\left.  +\frac{13357112}{35}\delta^{7} -\frac{280473801}{70}\delta^{8}+\frac{2769189970}{63}\delta^{9}
-\frac{788472308299}{1575}\delta^{10} \right), \\
|\psi^{\prime}_{310}(0)|^{2}&=
\frac{8}{729\pi a^{5}_{0}}\left(1-135 \delta^{2}+1215\delta^{3}-\frac{234495}{16}\delta^{4}+\frac{19793079}{80}\delta^{5} 
-\frac{752527017}{160}\delta^{6} \right. \nonumber \\
&\left. +\frac{53785680507}{560}\delta^{7} -\frac{74276863745607}{35840}\delta^{8}+\frac{334571340218103}{7168}\delta^{9}- \right. \nonumber \\
&\left. \frac{195134709782015919}{179200}\delta^{10} \right). 
\end{align} 
The large coefficients in these Taylor series raises again the concern on the corresponding convergence radii. The specific value of $\delta$ depends 
on the values of $M_{GB}$, the fine structure constant $\alpha_{D}$ and the dark matter mass $m$, and it is well possible to have 
large values of $\delta$. In order to have a reliable estimate in the large $\delta$ region we must resort again to analytic methods to 
enlarge the convergence radii, like the Pad\'{e} approximants  method. In Fig. \ref{WF2nl} we show result to different orders in the perturbative 
expansion as well as results using the $[5/5](\delta)$ Pad\'{e} approximant. 
It has been shown in \cite{Napsuciale:2020ehf} that variational methods underestimate the value of the squared wave function at the 
origin for the ground state when compared with the same order as calculated in our formalism. In Fig. \ref{WF2nl} we can see the 
effect of higher orders terms summed up by the Pad\'{e} approximants. The actual value of this observable decreases with 
increasing $\delta$ and vanishes at the critical value $\delta_{10}$. Similar results hold for $s$-wave higher states and the derivatives 
for $p$-waves. This could be important for darkonium phenomenology since, as mentioned above, for not so light mediator it is well 
possible that $\delta$ be in the critical region in whose case we will have weak transition matrix elements for darkonium. 

\section{Conclusions}
In this work we give a detailed phenomenological analysis of  the complete solution to the bound states of the Yukawa potential introduced in 
\cite{Napsuciale:2020ehf}. Eigenstates, $\psi_{nlm}$, and eigenvalues, $\epsilon_{nl}$, are obtained using the hidden supersymmetry of 
the system and a Taylor expansion of the Yukawa potential in terms of the parameter $\delta=a_{0}/D$.  
The solutions for the eigenvalues $\epsilon_{nl}$ are given as Taylor series in $\delta$ whose coefficients depend
on $n^{2}$ and $l(l+1)$. The eigenstates have a similar form to the solutions of the Coulomb potential but with the Laguerre associated 
polynomials, $L^{2l+1}_{n-l-1}(\rho)$, replaced with new polynomials of order $k$  in $\delta$, $N^{2l+1}_{n-l-1}(\rho,\delta,k)$, whose 
coefficients are polynomial in $\rho=2r/na_{0}$. We provide analytical expressions up to order $\delta^{10}$ for the eigenvalues and to 
order $\delta^{5}$ for the eigenstates but  both can be easily calculated to the desired order $k$. We wrote a {\it Mathematica} code,
publicly available, for this purpose \cite{Napsuciale:2021ma}.

Our results are in principle valid for small values of $\delta$ (large values of the screening) but since our procedure allows to easily 
calculate eigenvalues to higher orders, the convergence radii of the Taylor series can be enlarged up to the critical 
screening values using the Pad\'e approximants technique. This procedure permit us to calculate eigenvalues  with high precision, 
which we estimate from the properties of the Pad\'e approximants, in the whole range of values of $\delta$ where bound states exist. 
We find that, for a given $n$, states with higher $l$ are more energetic such that for $n\ge 4$ 
energy levels exhibit a cross-over phenomenon. The critical screening lengths, $\delta_{nl}$, can also be precisely calculated for 
every state and we give numerical results them up to $n=9$ using the expansion to very high order( from $\delta^{43}$ to $\delta^{53}$, 
depending of the state).

Similar considerations on the convergence radii apply to the eigenstates of the Yukawa potential. In this case, the information on the 
screening effects is contained in the new polynomials $N^{2l+1}_{n-l-1}(\rho,\delta,k)$. We provide analytical expressions up to $k=5$ 
for these polynomials. Higher order can be easily calculated but yields long expressions and are not shown here. We calculate the radial 
probabilities, finding that screening has sizable effects only for values of $\delta$ close to the critical ones. The convergence radii of 
the perturbative expansion of the new polynomials, can also be improved using the corresponding Pad\'e approximants. In this case, the 
coefficients of the rational functions entering the Pad\'e approximants are $\rho$-dependent which makes numerical computations more 
resource consuming. We provide results for $N=8$ (meaning a calculation to order $\delta ^{16}$) for the radial probabilities which allows 
us to reach screenings of the order $\delta\approx 3 \delta_{nl}/4$. For these values, a delocalization effect is already visible in the 
radial probabilities.

The squared absolute value at $r=0$ of the wave function for $s$-waves and its derivative for $p$-waves, enter the phenomenology of the 
decay and production of Yukawa bound states. In particular, for cold dark matter with a gauge structure and light dark mediators, Yukawa 
type darkonium could exist and these quantities are relevant to asses the possibilities of detecting darkonium in hadron colliders, direct 
or indirect detection experiments. We calculate $|\psi(0)|^{2}$ and $|\psi^{\prime}(0)|^{2}$ for $l=0$ and $l=1$ respectively, for the 
lowest energy levels and in the whole range of values of the screening lengths, using  the Pad\'e approximants technique.

Finally, our results can also yield some insight 
towards the construction of analytical solutions in closed form (not only as Taylor series). In this concern, it is interesting that 
in order to obtain our analytic solutions as Taylor series we require to perform a double factorization of
the $\tilde{H}^{(k)}_{l} $ superpartner Hamiltonians (see e.g. Eqs. (\ref{H2l},\ref{H2lbis})). Similar double factorization appears in 
supersymmetric quantum mechanics with supercharges realized with second order derivatives  \cite{Andrianov:1993md} and 
it is worthy to explore if such constructions and related developments 
\cite{Andrianov:1994aj} \cite{FernandezC.:2005bs} \cite{Correa:2015wxa} are useful in obtaining 
closed form solutions for the bound states of the Yukawa potential.  

\section*{Acknowledgments}

Dedicated to the memory of Arnulfo Zepeda, for his contribution to the development of the Mexican High Energy Physics community. 

\bibliographystyle{prsty}

\bibliography{Yukawa}

\section{Appendix}
The polynomials entering the solution of the Yukawa potential, up to order $k=5$,  for $n=1,2,3,4$ are given by   
\begin{align}
N^{1}_{0}(\rho,\delta,5)&=M^{5}_{1,0}(\frac{\rho}{2},\delta)= 
1- \left(\frac{3}{4}-\frac{\rho^2}{16}\right) \delta^2 
+ \left(\frac{11}{12}-\frac{\rho ^2}{24}-\frac{\rho^3}{144} \right) \delta^3  \nonumber \\
& - \left( \frac{395}{192} -\frac{\rho ^2}{96} -\frac{11 \rho^3}{1152} -\frac{\rho ^4}{384} \right) \delta^4 
  +\left(\frac{1647}{320} -\frac{\rho^2}{48} -\frac{5 \rho ^3}{384}-\frac{11 \rho ^4}{2560}  -\frac{7 \rho ^5}{14400} \right) \delta ^5 , \\
N^{1}_{1}(\rho,\delta,5)&=M^{5}_{2,0}(\rho,\delta)= 2-\rho  + \left( -24 +12 \rho+2 \rho ^2-\frac{\rho ^3}{2} \right)\delta ^2  \nonumber \\
 &+ \left(\frac{328}{3} -\frac{164 \rho }{3} -\frac{14 \rho^2}{3} +\frac{7 \rho ^3}{9} +  \frac{\rho ^4}{9}\right) \delta ^3  \nonumber \\
  &  +\left(-\frac{2500}{3}+\frac{1250 \rho}{3}-\frac{23 \rho ^2}{6}+\frac{95 \rho ^3}{36} +\frac{19 \rho ^4}{72}  -\frac{7\rho ^5}{48}\right) \delta ^4 
     \nonumber \\
&+ \left(\frac{34344}{5}-\frac{17172 \rho }{5} +\frac{124 \rho ^2}{3}-16 \rho^3 -\frac{31 \rho ^4}{15} 
+\frac{106 \rho ^5}{225} +\frac{53 \rho ^6}{900}\right) \delta ^5,  \\
N^{3}_{0}(\rho,\delta,5)&=M^{5}_{2,1}(\rho,\delta)= 1+ \left(-15+\frac{\rho ^2}{2}\right)\delta ^2 
+ \left(\frac{160}{3}-\rho ^2 -\frac{\rho^3}{9}\right) \delta ^3  \nonumber \\
&+  \left(-\frac{2225}{6}-\frac{11 \rho ^2}{4}+\frac{19 \rho ^3}{36}+\frac{7\rho ^4}{48}\right) \delta ^4 \nonumber \\
&+\left(\frac{14412}{5}+\frac{193 \rho ^2}{15}-\frac{23 \rho^3}{15}-\frac{7 \rho ^4}{10}-\frac{53 \rho ^5}{900}\right) \delta ^5 , \\
N^{1}_{2}(\rho,\delta,5)&=M^{5}_{3,0}(\frac{3\rho}{2},\delta)=  3-3 \rho +\frac{\rho ^2}{2}
+ \left(-\frac{729}{4}+\frac{729 \rho }{4}-\frac{243 \rho^2}{16}-\frac{135 \rho ^3}{16}+\frac{27 \rho ^4}{32}\right) \delta ^2  \nonumber \\
   &+ \left(\frac{7371}{4}-\frac{7371 \rho }{4}+\frac{459 \rho ^2}{2}+\frac{621 \rho ^3}{16}-\frac{27\rho ^4}{16}-\frac{9 \rho ^5}{32}\right) \delta ^3   
   \nonumber \\
   &+  \left(-\frac{1950075}{64}+\frac{1950075 \rho}{64}-\frac{677241 \rho ^2}{128}+\frac{20169 \rho ^3}{128}
   -\frac{2673 \rho ^4}{256}-\frac{243\rho ^5}{32}+\frac{405 \rho ^6}{512}\right) \delta ^4   \nonumber \\
   & +\left(\frac{171476109}{320} - \frac{171476109\rho }{320}+\frac{60293403 \rho ^2}{640}-\frac{379809 \rho ^3}{128}  \right. \nonumber \\
  & \left. +\frac{400221 \rho^4}{2560}+\frac{996057 \rho ^5}{12800}-\frac{63909 \rho ^6}{25600}-\frac{3159 \rho^7}{6400}\right) \delta ^5 ,  \\
N^{3}_{1}(\rho,\delta,5)&=M^{5}_{3,1}(\frac{3\rho}{2},\delta)= 4-\rho 
+ \left(-270+\frac{135 \rho }{2}+\frac{45 \rho ^2}{4}-\frac{27 \rho^3}{16}\right) \delta ^2  \nonumber \\
& + \left(2430-\frac{1215 \rho }{2}-54 \rho ^2  \frac{9 \rho ^3}{2}+\frac{9\rho ^4}{16}\right) \delta ^3  \nonumber \\
 &+\left( -\frac{307395}{8}+\frac{307395 \rho }{32}-243 \rho^2+\frac{567 \rho ^3}{8}+\frac{1053 \rho ^4}{128}-\frac{405 \rho ^5}{256}\right) \delta^4
\nonumber \\ 
&+\left(\frac{26354079}{40} -\frac{26354079 \rho }{160}+\frac{38637 \rho ^2}{10}-\frac{3888 \rho^3}{5}  \right. \nonumber \\
&\left. -\frac{34263 \rho ^4}{320}+\frac{105381 \rho ^5}{12800}+\frac{3159 \rho ^6}{3200}\right) \delta^5,  
\end{align}
\begin{align}
N^{5}_{0}(\rho,\delta,5)&=M^{5}_{3,2}(\frac{3\rho}{2},\delta)=  1+ \left(-\frac{189}{2}+\frac{27 \rho ^2}{16}\right)\delta^2
+ \left(\frac{1323}{2}-\frac{27 \rho ^2}{4}-\frac{9 \rho ^3}{16}\right)\delta^3 \nonumber \\
&+\left(-\frac{270459}{32}-\frac{729 \rho ^2}{8}+\frac{729 \rho ^3}{128}+\frac{405 \rho^4}{256}\right) \delta ^4 \nonumber \\
&+ \left(\frac{21437703}{160}+\frac{3645 \rho ^2}{4}-\frac{2187 \rho^3}{128}-\frac{7533 \rho ^4}{512}-\frac{3159 \rho ^5}{3200}\right)\delta ^5,  \\
N^{1}_{3}(\rho,\delta,5)&=M^{5}_{4,0}(2\rho,\delta)= 4-6 \rho +2 \rho ^2-\frac{\rho ^3}{6} \nonumber \\
&+ \left(-768+1152 \rho -320 \rho ^2-24 \rho ^3+12 \rho^4-\frac{2 \rho ^5}{3}\right)\delta ^2  \nonumber \\
&+ \left(\frac{41216}{3}-20608 \rho +\frac{18880 \rho^2}{3}-\frac{832 \rho ^3}{9}-\frac{256 \rho ^4}{3}+\frac{8 \rho ^5}{9}+\frac{8 \rho^6}{27}\right) \delta^3 
 \nonumber \\
&+ \left(-\frac{1195520}{3}+597760 \rho -\frac{607168 \rho^2}{3}+\frac{177568 \rho ^3}{9}-624 \rho ^4-\frac{800 \rho ^5}{9} \right.  \nonumber \\
&+\left. \frac{760 \rho ^6}{27}-\frac{13\rho ^7}{9}\right) \delta^4 
+ \left(\frac{61028352}{5} -\frac{91542528 \rho }{5}+\frac{93445888\rho ^2}{15}
 \right.\nonumber \\
&- \left. \frac{9352064 \rho ^3}{15}+\frac{107712 \rho ^4}{5}
+\frac{300608 \rho^5}{225}-\frac{90272 \rho ^6}{225} 
+\frac{8 \rho ^7}{75}+\frac{824 \rho ^8}{675}\right) \delta^5,  \\
N^{3}_{2}(\rho,\delta,5)&=M^{5}_{4,1}(2\rho,\delta)=  10-5 \rho +\frac{\rho ^2}{2}
+\left(-2040+1020 \rho -10 \rho ^2-30 \rho ^3+2 \rho^4\right) \delta^2 \nonumber \\
&+ \left(\frac{102400}{3}-\frac{51200 \rho }{3}+\frac{2720 \rho^2}{3}+\frac{2000 \rho ^3}{9}-\frac{40 \rho ^4}{9}-\frac{8 \rho ^5}{9}\right)\delta^3 
\nonumber \\
&+\left(-\frac{2894320}{3}+\frac{1447160 \rho }{3}-\frac{161900 \rho ^2}{3}+\frac{22480 \rho^3}{9}-\frac{20 \rho ^4}{9} \right.  \nonumber \\
&- \left. \frac{610 \rho ^5}{9}+\frac{13 \rho ^6}{3}\right)\delta^4
+\left(29133312-14566656 \rho +\frac{24671104 \rho ^2}{15} \right.  \nonumber \\
&-\left.\frac{202240 \rho ^3}{3} 
+\frac{544 \rho^4}{3}+\frac{48368 \rho ^5}{45}-\frac{56 \rho ^6}{5}-\frac{824 \rho ^7}{225}\right)\delta^5,  \\
N^{5}_{1}(\rho,\delta,5)&=M^{5}_{4,2}(2\rho,\delta)= 6-\rho +\left(-1512+252 \rho +36 \rho ^2-4 \rho ^3\right) \delta^2  \nonumber \\
+& \left(21504-3584 \rho-288 \rho ^2+16 \rho ^3+\frac{16 \rho ^4}{9}\right) \delta^3 \nonumber \\
+& \left(-536592+89432 \rho -4176 \rho^2+736 \rho ^3+\frac{620 \rho ^4}{9}-\frac{26 \rho ^5}{3}\right) \delta^4  \nonumber \\
  +& \left(\frac{77930496}{5}-\frac{12988416 \rho }{5}+94464 \rho ^2-12416 \rho ^3-\frac{4384 \rho^4}{3}  \right.
  +\left.\frac{1648 \rho ^5}{25} + \frac{1648 \rho ^6}{225}\right) \delta^5 ,  \\
N^{7}_{0}(\rho,\delta,5)&=M^{5}_{4,3}(2\rho,\delta)=   1+ \left(-360+4 \rho ^2\right) \delta ^2 
+  \left(4160-\frac{80 \rho ^2}{3}-\frac{16 \rho^3}{9}\right)\delta^3  \nonumber \\
&+  \left(-76400-\frac{2920 \rho ^2}{3}+\frac{280 \rho ^3}{9}+\frac{26 \rho^4}{3}\right) \delta^4  \nonumber \\
&+ \left(\frac{10350336}{5}+16576 \rho ^2-\frac{64 \rho ^3}{15}-\frac{656 \rho ^4}{5}-\frac{1648 \rho ^5}{225}\right) \delta ^5. 
\end{align}

\end{document}